\documentclass[preprint]{elsarticle}
\usepackage{lineno,hyperref}
\modulolinenumbers[1]
\usepackage[utf8]{inputenc}
\usepackage{amsmath}
\usepackage{xcolor}
\usepackage{multicol}
\usepackage{multirow}
\usepackage{amssymb}
\usepackage{comment}
\usepackage{caption}

\usepackage{graphicx}
\usepackage{subfigure}
\usepackage[autostyle]{csquotes}
\usepackage[american]{babel}
\usepackage[amssymb]{SIunits}
\usepackage{subfigure}
\usepackage{notoccite}
\usepackage{afterpage}
\usepackage{natbib}
\usepackage{epstopdf}
\usepackage{soul}
\journal{arXiv}

\begin{document}

\begin{frontmatter}

\title{Investigating the origin of cube texture during static recrystallization of fcc metals : A full field crystal plasticity-phase field study}

\author[a]{Supriyo Chakraborty}
\ead{chakraborty.84@osu.edu}
\author[a]{Chaitali S. Patil}
\ead{patil.152@osu.edu}
\author[a]{Yunzhi Wang}
\ead{wang.363@osu.edu}
\author[a]{Stephen R. Niezgoda\corref{correspondingauthor}}
\ead{niezgoda.6@osu.edu}
\address[a]{
  Department of Materials Science and Engineering, 
  The Ohio State University, Columbus, OH, USA
}

\cortext[correspondingauthor]{Corresponding author}

\begin{abstract}
The origin of cube recrystallization texture in medium to high stacking-fault energy fcc metals has been debated for almost 70 years. Despite numerous experimental and simulation studies, many issues regarding the nucleation and growth of cube grains remain unresolved. Here we apply a full field crystal plasticity model utilizing a dislocation density based constitutive theory to study the deformation and texture evolution in copper (Cu) under plane strain compression. Additionally, we use the phase field method, along with a stochastic nucleation model, for static recrystallization simulations. Simulation results show that the volume fraction of the cube component during deformation decreases with increasing strain. Although cube grains are not stable during plane strain compression, some of the non-cube grains rotate towards cube and develop narrow cube bands near the grain boundary region. With increasing deformation, the cube component accumulates dislocation density faster than other texture components. High stored energy in the cube regions leads to preferential nucleation of cube grains during static recrystallization. These cube nuclei originate from the intergranular cube bands.  Although the cube component has a clear nucleation advantage, none of the texture component appears to have a growth advantage. Instead, simulation results show that heterogeneous distribution of nuclei has a profound influence on the resulting grain size distribution. During recrystallization, a significant increase in cube volume fraction is observed mainly due to high nucleation frequency of cube grains. 
\end{abstract}


\end{frontmatter}

\section{Introduction}


 Crystallographic texture plays an important
role in wide range of engineering applications \cite{kocks1998,suwas2014}. For example, a large fraction of cube oriented (\{100\}$\langle001\rangle$) grains in aluminum sheets lead to plastic anisotropy during deep drawing processes. Therefore, to maintain uniform material flow, sharp cube texture is undesirable in aluminum sheet \cite{tucker1961,engler1996}. In case of electrical steels, $\langle001\rangle$ direction provides easy magnetization, hence, Goss oriented (\{110\}$\langle001\rangle$) grains are preferred to minimize magnetic core losses \cite{matsuo1989,dorner2007}.  
 Similarly, nickel substrates with strong cube textures are favored for epitaxial growth of high temperature superconductors \cite{bhattacharjee2007nickel}. 
  
 Medium to high stacking fault energy fcc metals develop a pronounced cube texture after cold rolling followed by high temperature annealing \cite{merlini1953,Ridha,hjelen1991,duggan1993}. During this annealing treatment, ``strain free'' grains nucleate and grow; consuming the deformed microstructure. This process of nucleation and growth of new grains is commonly known as primary or static recrystallization \cite{humphreys2012}. Unless subject to additional processing, cube texture developed during static recrystallization typically prevails in the final product. 
 Hence, understanding the deformation process during cold rolling and nucleation and growth processes during recrystallization is the key to control the cube texture in fcc metals.
  
Although the observation of a strong cube texture was universal, its origin has been strongly debated over the last 70 years. Two major competing theories, i.e. ``oriented  nucleation'' \cite{Ridha,hjelen1991,necker1970,Dillamore,samajdar1995,doherty1998cube,samajdar1999} and  ``oriented  growth'' \cite{beck1952,liebmann1956,lucke1974} have been promoted and vigorously debated as potential mechanisms for cube texture formation. Proponents of the oriented nucleation theory argued that cube oriented grains nucleate in much higher frequency than would be expected in a random grain structure. Whereas, proponents of oriented growth theory suggested that cube oriented grains preferentially grow in the deformed matrix and become larger than other grains. 

In support of oriented nucleation, Dillamore and Katoh \cite{Dillamore} suggested that cube oriented transition bands, which are stable along the normal direction (ND), form in between two diverging deformation bands. As these cube transition bands are located in regions of large lattice curvature, Dillamore and Katoh further claimed that cube transition bands are the preferred sites for nucleation. Although nucleation from cube oriented transition bands were observed in experimental studies \citep{Ridha,sindel1970}, the stability of the cube orientation along ND rotation was questioned by Driver and coworkers \cite{akef1991,basson2000} as well as by Wert et al. \cite{wert1997}. On the other hand, Doherty and Samajdar \cite{samajdar1995,samajdar1999,doherty1997} reported that cube nuclei originate from the deformed cube bands or debris of initial cube grains, therefore, a large fraction of cube grains in the undeformed material will lead to more cube grains following recrystallization. However, initial cube grains are not stable in plane strain compression \cite{akef1991,basson2000,liu1998,raabe2004} which poses a challenge for the hypothesis that nucleation occurs from the deformed cube bands.

Another popular hypothesis for the preferential nucleation of cube grains is that cube oriented subgrains recover rapidly due to the presence of dislocation pairs with orthogonal Burgers vectors \cite{Ridha}. Although this hypothesis was supported in previous studies \cite{sindel1970,doherty1997,samajdar1998}, direct evidence of such recovery mechanism is still lacking. Additionally, to the best of our knowledge no previous experiment or simulation studies investigated whether the formation of orthogonal dislocation pairs are unique to the cube oriented subgrains. 

On the other hand, the idea of oriented growth originated from the observation that certain grain boundaries (e.g. 40\degree about $\langle1 1 1\rangle$ axis in fcc metals) can have higher mobility relative to other grain boundaries \cite{beck1952,liebmann1956,lucke1974}. In the case of cube texture development in fcc metals, oriented growth became popular due to a near  40\degree-$\langle1 1 1\rangle$ orientation relationship between the ideal cube orientation and the $S$ orientation (\{1 2 3\}$\langle6 3 4\rangle$). Although this special orientation relationship for the cube grains may exist in some selected areas of deformed matrix, it is extremely difficult to maintain this relationship over a large area. This is partially due to the fact that the S orientation is not the only stable texture component in plane strain compression. Copper (\{1 1 2\}$\langle1 1 1\rangle$) and brass (\{1 1 0\}$\langle1 1 2\rangle$) texture components are also present at large volume fractions in deformed microstructure \cite{kocks1998,HIRSCH1988}. Both copper and brass do not possess the  40\degree-$\langle1 1 1\rangle$ orientation relationship with the cube component. Therefore, other mechanisms of preferential growth such as micro-growth selection \cite{duggan1993}, orientation pinning \cite{jensen1995,dorte} have been proposed over the years. However, these mechanisms also rely on the misorientation dependent grain boundary mobility of cube recrystallized grains. Nes and coworkers \cite{solberg,hjelen1991} discussed the effect of inhomogenous nucleation and impingement of recrystallized grains on the cube texture development. According to these authors, cube grains which are isolated from other nuclei can grow large due to lack of impingement. Although inhomogenous nucleation has been commonly observed during recrystallization, only a few studies have investigated its effect on texture development \citep{vandermeer1995,jensen1997}. 

Despite the large number of experimental and simulation studies, many issues regarding the origin of the cube texture remain unresolved. In this study, we focus on the following questions: 
\begin{enumerate}
\item What is the role of initial cube grains in nucleating cube oriented recrystallized grains? 
\item How does the stored energy evolve in different texture components?
\item Does the cube component has either a nucleation or growth advantage over other texture components?
\item Does nucleation of cube grains follow the Dillamore and Katoh mechanism?
\item What is the role of heterogeneous nucleation during recrystallization?
\item Why does the intensity of cube component increases sharply after recrystallization?
\end{enumerate}
In order to completely resolve these issues, knowledge of local orientation changes, slip system activity and dislocation density evolution in the bulk of a three-dimensional sample is necessary. Even with the recent advances in 3D experimental techniques such as High Energy Diffraction Microscopy \cite{bernier2011,chatterjee2016,wang2017,tayon2019}, it is still largely impossible to collect all the necessary information in-situ. Therefore, we must rely on physics-based simulation techniques to explore the long standing problem of cube texture development. 

Previous simulation studies were mainly conducted to explore recrystallization kinetics \cite{marthinsen,srolovitz,raabe2000,takaki,abrivard2012,moelans1,chenLQ}. However, only few of them focussed on cube texture development in fcc metals \cite{alvi2008,brahme2008,field,zecevic2019}. Alvi et al. \cite{alvi2008} and Brahme et al. \cite{brahme2008} explored recrystallization texture development following hot deformation. On the other hand, the work of Adam et al. \cite{field} primarily focussed on overall texture development, hence, emphasis on the origin of cube texture was limited. Recently, Zecevic et al. \cite{zecevic2019} used visco-plastic self-consistent (VPSC) model to study the oriented nucleation of cube grains, however, we believe a full-field model is necessary to truly capture the effect of local heterogeneities on both nucleation and growth of recrystallized grains. In this work, we use a full-field crystal plasticity model coupled with a dislocation density based constitutive theory to study the deformation process in polycrystalline copper under plane strain compression. A stochastic nucleation model is combined with phase field model to study the subsequent texture evolution during static recrystallization.

\section{Modeling framework}
\label{model}
\subsection{FFT based crystal plasticity model}

Moulinec and Suquet  \cite{moulinec} proposed a micromechanical solver utilizing the Green's function method and fast Fourier transform (FFT) approach. Later, it was extended by Lebensohn et al. \cite{lebensohn1} to formulate an elasto-viscoplastic fast Fourier transform (EVP-FFT) based crystal plasticity model. Recently, FFT based crystal plasticity solvers have emerged as robust tools for studying microstructure based mechanical problems. Previous studies using FFT based crystal plasticity models have been successfully employed to predict the local stress state during twinning in hcp crystals \cite{arul1,arul2}, stress and strain partitioning in dual phase steel \citep{tasan,diehl2017} and inter and intragranular misorientation 
changes during deformation \citep{lebensohn3,connor}. Additionally, FFT based crystal plasticity has been successfully coupled with phase field modeling to study static recrystallization \citep{chenLQ} and dynamic recrystallization
\citep{zhao1,zhao2}. More details on the numerical formulation of FFT based crystal plasticity models can be found in the recent review article of  Lebensohn and Rollett \cite{lebensohn_rollett}. In the present work we utilize the  EVP-FFT framework of Lebensohn et al. \cite{lebensohn1} to simulate the mechanical response of polycrystalline copper during room temperature deformation. EVP-FFT uses an Euler implicit time discretization scheme for the evolution of stress at
a material point x which is written as,
\begin{align}
\sigma^{t+\Delta t}(x) = C(x):\epsilon^{e,t+\Delta t}(x) = C(x):(\epsilon^{t+\Delta t}(x) - \epsilon^{p,t}(x) - \dot\epsilon^{p,t+\Delta t}(x)\Delta t)
\end{align}
where $\sigma(x)$ is the Cauchy stress tensor, $C(x)$ is the stiffness tensor and $\epsilon(x)$ is the strain tensor which is further decomposed into elastic ($\epsilon^{e}(x)$) and plastic ($\epsilon^{p}(x)$) components. The plastic strain rate $\dot\epsilon^{p,t+\Delta t}(x)$ can be calculated using a single crystal constitutive model.

 \subsubsection{Single crystal constitutive model}
Both nucleation
and growth of the recrystallized grains depend on the amount of stored energy in the deformed material; which in turn is directly related to the elastic strain energy of the dislocations. Therefore, to successfully capture all the relevant physics behind deformation and recrystallization a dislocation density based constitutive model is required.
Additionally, experimental work of Doherty \citep{doherty1997} has shown that the initial grain size also has a significant influence on the kinetics and texture evolution during static recrystallization. Therefore, a nonlocal constitutive model is required to capture the effect of grain size on the stored energy distribution and texture evolution during plastic deformation. For this purpose, we mainly followed the framework of Ma, Raabe, and Roters \cite{ma1,ma2} to formulate a dislocation density based constitutive model for the room temperature deformation of copper.   

At the highest level, slip activity is described by the Orowan equation,
\begin{align}
 \dot{\gamma^{\alpha}} = \rho_{m}^{\alpha}bv^{\alpha}
\end{align}
where $\dot{\gamma^{\alpha}}$ is the shear strain rate, $\rho_m^{\alpha}$ is the mobile dislocation density, $b$ is the Burgers vector and $v^{\alpha}$ is the velocity of the mobile dislocations for slip system $\alpha$. Dislocation motion is assumed to occur via thermally
activated jumps of mobile dislocations temporarily pinned by short range obstacles (e.g. forest dislocations) on the slip plane \citep{conrad,kocks}. Then the velocity $v^{\alpha}$ can be expressed as a thermally activated rate equation,
\begin{align}
 v_{slip}^{\alpha} = \dfrac{1}{2}\lambda\nu \exp{\left(-\dfrac{Q_{slip}}{KT}\left(1 - \dfrac{\tau_{app}^{\alpha} - \tau_{pass}^{\alpha}}{\tau_0 + \tau_{cut}^{\alpha}}\right)\right)}
 \end{align}
where $\lambda$
is the jump distance, $\nu$
is the jump frequency,
$Q_{slip}$ is the activation energy required to overcome the obstacle,
$\tau_{app}^{\alpha}$ is the resolved shear stress for slip system $\alpha$.
$\tau_{pass}^{\alpha}$ is the back stress from parallel dislocations on the same slip system, $\tau_{cut}^{\alpha}$ is the stress required to overcome the obstacle without any thermal activation and $\tau_0$ is the stress required to overcome lattice friction (Peierls barrier). Lattice friction stress was not considered in the original works of Ma, Raabe and Roters \cite{ma1,ma2}, however, we found that a small value of $\tau_0$ improves the numerical stability and accuracy of the predicted initial hardening rate with a low initial dislocation density. Expressions for $\lambda$, $\tau_{pass}^{\alpha}$ and $\tau_{cut}^{\alpha}$ can be found in \cite{ma1,ma2}. 

Evolution of immobile dislocation density is written as,
\begin{align}
\dot{\rho_{im}}^{\alpha} = C_4|\dot{\gamma^{\alpha}}|\sqrt{\rho_F^{\alpha}} + C_5|\dot{\gamma^{\alpha}}|d_{dipole}^{\alpha}\rho_m^{\alpha} - C_6|\dot{\gamma^{\alpha}}|\rho_{im}^{\alpha} - C_7d_c(\rho_{im}^{\alpha})^2
\label{dd_evolution_eqn}
\end{align}
where $\rho_{im}^{\alpha}$ is the immobile dislocation density and $\rho_F^{\alpha}$ is the forest dislocation density for the slip system $\alpha$. $d_{dipole}^{\alpha}$ is the dipole formation distance between two mobile dislocations. $d_c$ is the critical dipole separation distance below which a dipole becomes unstable and the dislocations in the dipole will spontaneously annihilate \cite{mughrabi}. $C_4, C_5, C_6$  and $C_7$ are fitting parameters and their values are given in Table \ref{fitting_para}.

The first two terms in Eq. \ref{dd_evolution_eqn} represent the generation rate of immobile dislocations due to lock formation and dipole formation, respectively. The third term represents the rate of annihilation of immobile dislocations due to interactions with mobile dislocations. 
We neglected dislocation climb controlled recovery as it is extremely difficult to activate at room temperature due to significantly large activation barrier for lattice diffusion in copper. 
Instead, we introduced an additional term for spontaneous annihilation of dislocations in dipoles when the dipole separation distance reaches a critical value. Although \cite{ma1,ma2} did not consider this reaction in their model, Essmann and Mughrabi \cite{mughrabi} has emphasized its importance. Recently, Kords \cite{kords} also considered the spontaneous annihilation of dipoles in their constitutive model. \cite{mughrabi} has reported that the critical dipole separation distance ($d_c$) for edge dislocations is around 1.6 nm for copper.

We followed the approach of Dai \cite{dai}, Arsenlis and Parks \cite{arsenlis2} to calculate the GND density from the strain gradient. Dai \cite{dai} decomposed the total GND density into screw and edge components which in small strain
description takes the following form,
\begin{align}
\dot{\rho^{\alpha}_{G(s)}} = - \dfrac{1}{b}\vec{\nabla} \dot{\gamma^{\alpha}}.\vec{t^{\alpha}}
\\  
\dot{\rho^{\alpha}_{G(e1)}} =  \dfrac{1}{b}\vec{\nabla} \dot{\gamma^{\alpha}}.\vec{d^{\alpha}}
\\
\dot{\rho^{\alpha}_{G(e2)}} = 0
\end{align}
where $\rho^{\alpha}_{G(s)}$ represents a set of screw dislocations for which line direction is along the slip direction $\vec{d^{\alpha}}$, $\rho^{\alpha}_{G(e1)}$ represents a set of edge dislocations for which line direction is along $\vec{t^{\alpha}}$ and $\rho^{\alpha}_{G(e2)}$ is a set of edge dislocations for which line direction is along the normal to the slip plane $\vec{n^{\alpha}}$.
The length scale associated with the gradient ($\Delta$x) provides a physical length scale to capture the size effect in materials.

Interaction between dislocations plays an important role in stage II strain hardening \cite{bulatov}, therefore, the strength parameters need to be carefully chosen. Recently, Kubin et al. \cite{kubin2008}, Madec and Kubin \cite{madec2017} and Martinez et al. \cite{martinez2008} conducted discrete dislocation dynamics simulations at different temperatures and
showed that the interaction between two dislocations with same Burgers vector on cross slip planes
(collinear interaction) has the highest interaction strength.
Among the possible non-coplanar
interactions the Hirth lock is the weakest and  Lomer-Cottrell and glissile junction strengths fall in
between the collinear interaction and Hirth lock strength. Therefore, we used the differential interaction strengths for copper calculated by Madec and Kubin \cite{madec2017} in our constitutive model. Values of all the interaction parameters are given in Table \ref{interaction}.

\subsection{Nucleation model}
\label{nucleation_model}
During static recrystallization nucleation events occur at length and time scales significantly smaller than what is accessible by either crystal plasticity or phase field. Therefore, a stochastic nucleation model is necessary in which the probability of nucleation relies on parameters which can be defined in our current modeling framework.  Following the work of Niezgoda et al. \cite{niezgoda} we used Weibull probability distribution function to draw a random variate $k(x)$  for each material point \lq x\rq. The probability distribution function reads as,
\begin{align}
    f(k(x);k_c,\eta) = \dfrac{\eta}{k_c}\left(\dfrac{k(x)}{k_c}\right)^{\eta -1}exp^{\left(\dfrac{-k(x)}{k_c}\right)^\eta} \forall \hspace{3pt}  k(x)\geq 0
\end{align}
where $k_c$ and $\eta$ are the scale parameter and shape parameter of the Weibull distribution, respectively. To ensure random nucleation the variates were generated using a seed value which changes with the computer clock time. 

Based on experimental observations \cite{Ridha} and theoretical arguments \cite{doherty1997current,szpunar,holm2003,rollett}, it is understood that during static recrystallization nucleation events are most likely to occur at high stored energy locations where chances of accelerated recovery are high due to fast annihilation of dislocations and abnormal subgrain growth. Therefore, in this model, nucleation will occur when the total dislocation density $\rho_{tot}(x)$ exceeds the value of drawn variate $k(x)$. Therefore, nucleation of recrystallized grains can occur anywhere in the microstructure if it satisfies the above condition. The total dislocation density is calculated as,
\begin{align}
    \rho_{tot}(x) = \sum^N_{\alpha = 1} \rho^{\alpha}_s(x) + \rho^{\alpha}_m(x) + \sqrt{(\rho^{\alpha}_{G(s)}(x))^2 + (\rho^{\alpha}_{G(e1)}(x))^2}
\end{align}
where N is the number of independent slip systems.

A nucleus with radius of three grid points (3$\Delta$x) was placed at each location where nucleation condition was satisfied. As we mentioned above, recrystallized grains nucleate from the deformed substructure due to rapid recovery, not by random atomic fluctuations in the classical nucleation  sense. Hence, recrystallized nuclei generally inherit the crystallographic orientation of the deformed state. Therefore, we assigned the local crystallographic orientation of the deformed state to each recrystallized nucleus. Moreover, dislocation density in the recrystallized nuclei was set to the initial value to make the recrystallized nuclei \enquote{strain free}.

The stochastic nucleation model determines only the number of nuclei appearing throughout the simulation volume. However, for realistic modeling of recrystallization a time scale is required to truly capture the effect of nucleation on recrystallization kinetics. Here we introduce a nucleation attempt frequency $u^{nuc}$ which will control the timescale of nucleation. In simple terms, variates from the Weibull distribution will be drawn $u^{nuc}$ number of times in every second. Hence, $k_c$, $\eta$ and $u^{nuc}$ all together will determine the actual nucleation rate during recrystallization. Values for the nucleation model parameters are listed in Table \ref{nucleation_para}.

\subsection{Phase Field Formulation}
During static recrystallization newly formed grains grow in the deformed matrix due to the driving force associated with the reduction of stored energy. As recrystallized grains have lower stored energy than deformed grains, stored energy difference between recrystallized and deformed grains act as the driving force for grain growth. Previously, Moelans et al. \cite{moelans1} and Chen et al. \cite{chenLQ} used continuum field phase field model to simulate static recrystallization. In these models, stored energy of the recrystallized grains was ignored in the free energy formulation. However, in our crystal plasticity simulations dislocation density can vary orders of magnitude in the deformed grains due to plastic heterogeneities (see Figure \ref{dislocation_spatial}). The driving force for grain growth will be reduced in the regions where dislocation density remain low. Hence, to precisely measure the stored energy difference, we considered the stored energy of both recrystallized and deformed grains in our free energy formulation. Moreover, \cite{moelans1,chenLQ} used two separate order parameters to define deformed and recrystallized grains. He we differentiate  deformed and recrystallized regions through the local stored energy. Therefore, we adopted a single set of order parameters which will also increase the computational efficiency.

The phase-field formulation is similar to the continuum field models of Chen and Wang\cite{chenLQ1}, Fan and Chen \cite{chenLQ2}. The stored energy due to dislocation content is added as bulk energy in the free energy functional. The free energy functional in our model takes the following form,
\begin{align}
 F = \int ( m f_{0}(\phi_{1} , \phi_{2}, ... \phi_{N}) + f_{gr}(\nabla \phi_{1}, \nabla \phi _{2}, ... \nabla \phi_{N}) + f_{st})dV
\end{align}
where $\phi_{1},\phi_{2}...\phi_{N}$ are the phase field order parameters, $f_{0}$ is a well type potential, m is the height of the well, $f_{gr}$ is the gradient free energy density and $f_{st}$ is the stored energy density. 

For the term $f_{0}$ we used the same formulation used in previous studies by Moelans et al. \cite{moelans2008},
\begin{align}
 f_{0} = \sum_{i=1}^N (\dfrac{\phi_{i}^4}{4} - \dfrac{\phi_{i}^2}{2}) + \Gamma \sum_{i=1}^N \sum_{j>i}^N \phi_{i}^2\phi_{j}^2 + \dfrac{1}{4}
\end{align}
The gradient free energy density takes the following form,
\begin{align}
 f_{gr} = \sum_{i=1}^N \dfrac{\kappa}{2}(\nabla \phi_{i})^2
\end{align}
where $\kappa$ is the gradient energy coefficient. $f_{st}$ is written as the summation of stored energy densities associated with all the order parameters,  
\begin{align}
f_{st} = \sum_{i=1}^N\dfrac{h(\phi_{i})}{\sum_{i=1}^N h(\phi_{i})} f_{i}^d
\label{stored_equation}
\end{align}
where $f_{i}^d$ is the stored energy density in the \enquote*{i}th grain,
\begin{align}
 f_{i}^d = \dfrac{1}{2}\mu b^2 \rho_{i}
\end{align}
where $\rho_{i}$ is the dislocation density in the \lq i'th grain. $\dfrac{h(\phi_{i})}{\sum_{i=1}^N h(\phi_{i})}$ is an interpolation function which is 
formulated in a thermodynamically consistent manner \cite{moelans3}. Here we used a fifth order polynomial function for $h(\phi_{i})$,
\begin{align}
 h(\phi_{i}) = \phi_{i}^3(6\phi_{i}^2 - 15\phi_{i} + 10)
\end{align}
We used the Allen-Cahn equation for the evolution of order parameters,
\begin{align}
\dfrac{\partial \phi_{i}}{\partial t} 
= -L\left(m(\phi^{3}_{i} - \phi_{i} + 2\Gamma\phi_{i}\sum^{N}_{j\neq i}\phi^2_{j}) - \kappa\nabla^2\phi_{i} + \dfrac{h^{'}(\phi_{i})\sum^N_{j\neq i}h(\phi_j)\mu b^2(\rho_i - \rho_j)}{2(\sum^{N}_{i=1}h(\phi_{i}))^2}   \right)
\label{AC}
\end{align}
where L is a kinetic parameter which is written as $L = \dfrac{M_{gb}E_{gb}}{\kappa}$ where $M_{gb}$ is the mobility of grain boundary and $E_{gb}$ is the grain boundary energy. $h^{'}(\phi_{i})$ is the derivative of $h(\phi_i)$ with respect to $\phi_i$. 
The driving force in Eq. \ref{AC} (last term in right hand side) becomes zero in the grain interior because $h^{'}(\phi_{i})$ and $h(\phi_j)$ takes the value of zero in the grain interior. Therefore, the driving force for grain growth acts only at the grain boundary. On the other hand, we want to emphasize that nucleation of new recrystallized grains can still occur both at grain boundary and grain interior.

The stored energy of a grain in Eq. \ref{stored_equation} can be calculated either by taking an average dislocation density for the whole grain or by taking a local average of dislocation density across the diffuse interface as the order parameters only evolve near the diffuse interface. For the present work, as dislocation density distribution is highly heterogeneous in the deformed matrix, all stored energy calculations were done locally near the diffuse interface to capture the anisotropic growth of recrystallized grains.

To solve Eq. \ref{AC} at location $(x,y,z)$ we calculate
the stored energy densities by taking an average dislocation density from a small cubic
box spanning over $(x \pm \delta),(y \pm \delta),(z \pm \delta)$ where $\delta$ depends on the number of grid points in the diffuse interface. A similar but slightly different averaging technique has been adopted by \cite{takaki} in their recrystallization model.  

\section{Results and Discussions}

In this section, we will first present texture and dislocation density evolution during plane strain compression. Next, we will discuss simulation results in terms of nucleation, growth and texture evolution during static recrystallization. All the texture analysis was done using the open source MTEX-4.3.2 software \cite{bachmann2010}. All the texture components were measured within 15 degrees from their ideal orientation and triclinic sample symmetry was assumed. Euler angles of the texture components found in plane strain compression and static recrystallization are listed in \ref{Euler_texture_comp}.  
\begin{table}[!htbp]
    \caption{Euler angles of the main texture components \cite{kocks1998}.}
\label{Euler_texture_comp}
 \begin{tabular}{l l}
 \hline
 component & Euler angles (Bunge convention)\\
 \hline
     cube    & 0\degree, 0\degree, 0\degree  \\
     copper    & 90\degree, 35\degree, 45\degree \\
     S & 59\degree, 37\degree, 63\degree \\
     brass & 35\degree, 45\degree, 0\degree \\
     Goss & 0\degree, 45\degree, 0\degree \\
     \hline
    \end{tabular}
\end{table}

\begin{figure}[!htbp]
    \centering
    \includegraphics[height=0.5\linewidth]{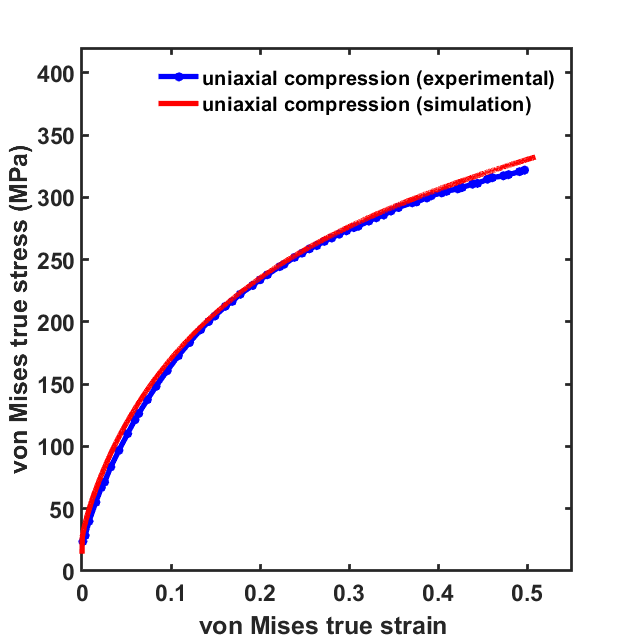}
    \caption{Stress-strain plots for experimental uniaxial compression test at room temperature \cite{bronkhorst} (blue line), simulated uniaxial compression test at 300 K (red line). For both experiment and simulation, applied strain rate was 0.001/s.}
    \label{SS}
\end{figure}

Crystal plasticity model parameters were calibrated with an uniaxial compression test data available in literature \cite{bronkhorst}. The uniaxial compression simulation was conducted with a strain rate of 0.001/s at room temperature. The experimental along with the calibrated stress-strain curves are shown in Figure \ref{SS}. Fitting parameters obtained from calibration are listed in Table \ref{fitting_para}.

\subsection{plane strain compression simulation}
\label{deformation}
Plane strain compression simulations were performed using the fitted model parameters with an applied strain rate of 0.001/s.
\begin{table}[!htbp]

\caption{Parameters of dislocation density based constitutive model for copper.}
\label{fitting_para}
 \begin{tabular}{l l}
 \hline
     parameter & value\\
\hline
     $Q_{slip}$ & $2.0\times10^{-19}\quad J$\\
     $\rho^{\alpha,initial}_{im}$ & $5.0\times10^{11}\quad m^{-2}$\\ 
     $\nu$ & $10^{10} s^{-1}$\\
     $C_1$ & 0.51\\
     $C_2$ & 3.0\\
     $C_3$ & 3.0\\
     $C_4$ & $8.0\times10^{8}\quad m^{-1}$\\
     $C_5$ & $1.1\times10^{12}\quad m^{-1}$\\
     $C_6$ & $15.0$\\
     $C_7$ & $9.0\times10^{-9} \quad ms^{-1}$\\
     $\Delta x$ & $3.0\times10^{-6}\quad m$\\
\hline
\end{tabular} 
\end{table}
\begin{table}[!htbp]

\caption{Dislocation interaction strength parameters for copper \cite{kubin2008,madec2017}.}
\label{interaction}
 \begin{tabular}{ll}
 \hline
Interaction & $\chi_{\alpha\beta}$ \\
\hline
self & 0.13\\
coplanar & 0.13\\
Hirth lock & 0.05\\
glissile junction & 0.13\\
collinear & 0.72\\
Lomer-Cottrell lock & 0.18\\
\hline
\end{tabular}
\end{table}
In this regard, we used two representative volume elements (RVE) where each of them has total 128$\times$128$\times$128 grid points but a different initial texture. The first RVE has total 491 grains with very low volume fraction of cube oriented grains, we call it \lq RVE1'  and the second RVE has total 488 grains with higher volume fraction of cube oriented grains, we call it \lq RVE2'. The purpose of using two different RVEs is to understand the effect of microstructural variability and initial cube texture on the deformation and subsequent recrystallization. The average grain size in both RVE1 and RVE2 is 60.5 $\mu$m. 

\subsubsection{Deformation texture evolution}
\begin{figure}[!htbp]
   \centering
    \subfigure[]{
    \includegraphics[width=0.5\linewidth]{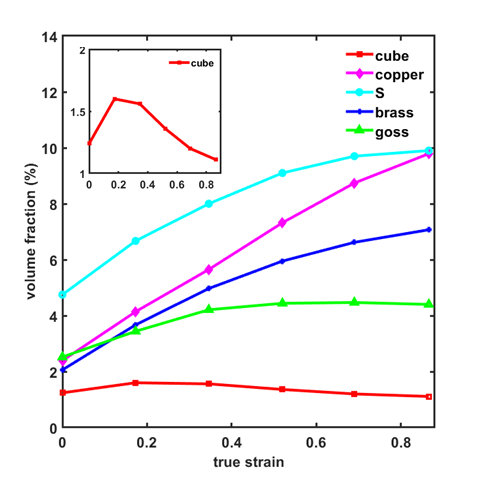}
\label{texture_RVE1}
}

\subfigure[]{
    \includegraphics[width=0.5\linewidth]{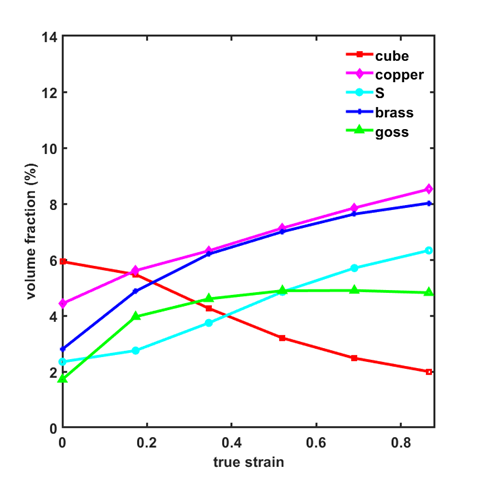}
\label{texture_RVE2}
}
   \caption{Development of various rolling texture components with increasing strain under plane strain compression for (a) RVE1 and (b) RVE2. Inset in (a) shows the magnified view of the cube component in RVE1.} 
    \label{texture_evolution}
\end{figure}

Figure \ref{texture_evolution} shows the evolution of main rolling texture components for both RVE1 and RVE2 with increasing levels of deformation. For both microstructures, volume fraction of the copper, S and brass components increase with the increasing deformation (although the actual fraction varies because of their different initial volume fractions). This type of texture evolution is well known for cold rolling of medium to high SFE fcc metals and is commonly  known as \lq copper type' texture \cite{kocks1998}. Volume fraction of the Goss component first increases and then saturates after a small deformation. A close inspection of the cube component in RVE1 (as shown in the inset of Figure \ref{texture_RVE1}) reveals that volume fraction of the cube component first increases by a small amount and then decreases with increasing strain. On the other hand, in RVE2 (as shown in Figure \ref{texture_RVE2}) volume fraction of the cube component  decreases steadily with increasing strain. Therefore, at a first glance behavior of the cube component seems contrasting in these two cases. 

From previous experimental and simulation studies on fcc metals it is known that cube oriented grains are not stable under plane strain compression \cite{akef1991,basson2000,liu1998,raabe2004}. Hence, a  decrease in the cube volume fraction in both Figure \ref{texture_RVE1} and \ref{texture_RVE2} is expected. In contrast, an initial increase of cube component in Figure \ref{texture_RVE1} signifies that some non-cube orientated regions rotated towards cube orientation during deformation. This small increase in cube fraction can be noticed only if initial volume fraction of the cube component is low, as in the case for RVE1. On the other hand, if initial cube volume fraction is sufficiently high which is the case for RVE2 then any increase in cube volume fraction due to rotation of non-cube grains will be small compared to the overall decrease in cube fraction. 
 
\begin{figure}[!htbp]
    \centering
     \subfigure[$\epsilon_{vM}$ = 0.0]{
    \includegraphics[height=0.4\linewidth]{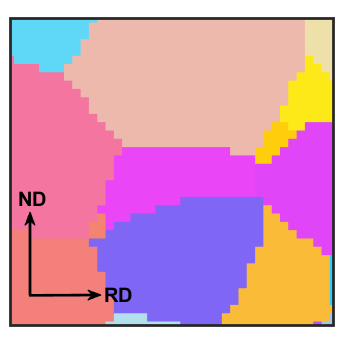}
    }
    \subfigure[$\epsilon_{vM}$ = 0.17]{
    \includegraphics[height=0.4\linewidth]{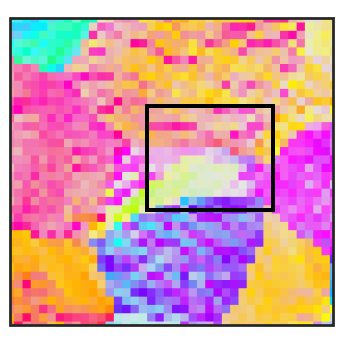}
    }
    \subfigure[$\epsilon_{vM}$ = 0.52]{
    \includegraphics[height=0.4\linewidth]{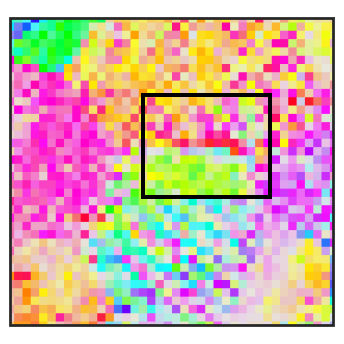}
    }
    \subfigure[$\epsilon_{vM}$ = 0.87]{
    \includegraphics[height=0.4\linewidth]{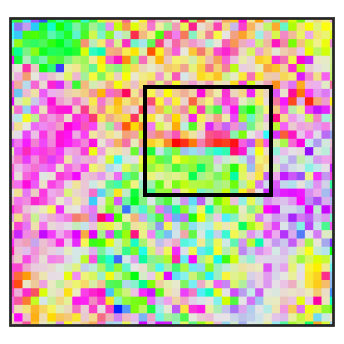}
    \label{TB_50}
    }
     
    \caption{Orientation maps showing the development of a cube oriented region along the grain boundary between two non-cube oriented grains. (a) shows the orientation map for the undeformed state where ND and RD arrows in (a) show the normal direction and rolling direction, respectively. Whereas (b), (c) and (d) show the orientation maps of the same region at increasing levels of deformation. The black box in (b), (c) and (d) highlight the location where cube oriented region forms.[0 0 1] inverse pole figure color code is used to represent the orientations.}
    \label{TB_band}
\end{figure}

\begin{figure}
    \centering
    \includegraphics[height=0.5\linewidth]{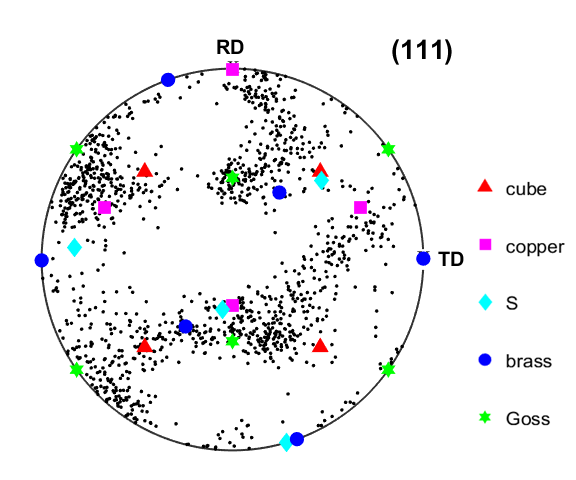}
    \caption{\{1 1 1\} pole figure showing the orientations (smaller black dots) in the region highlighted with the black box in Figure \ref{TB_50}. Ideal orientations of cube, copper, S, brass and Goss components are also shown with colored markers.}
    \label{TB_polefig}
\end{figure}

 After 50\% rolling reduction ($\epsilon_{vM}$ = 0.87) the total cube volume fraction retained in RVE1 is 1.1\% of which only 18\% (0.2\% overall fraction) belongs to the grains which were initially cube oriented. Similarly, total cube volume fraction retained in RVE2 is 2.0\% of which 45\% (0.9\% overall) belongs to the grains which were initially cube oriented. Therefore, cube orientations which are originated from non-cube grains dominate the cube texture component at larger strain. One such cube oriented region is shown in Figure \ref{TB_band}. This cube oriented region is formed as a narrow band along the grain boundary between two non-cube oriented grains. With increasing strain these two grains rotate towards copper and Goss orientations and the cube band forms in between them. In Figure \ref{TB_polefig}, the \{1 1 1\} pole figure shows the orientations in the region highlighted with the black box in Figure \ref{TB_50}. Clusters of orientation near the ideal copper and Goss orientation are clearly visible in the pole figure. Additionally, few orientations near the ideal cube orientation can also be observed.  
 
\subsubsection{Dislocation density distribution}

\begin{figure}[!htbp]
    \centering
\subfigure[]{
    \includegraphics[height=0.5\linewidth]{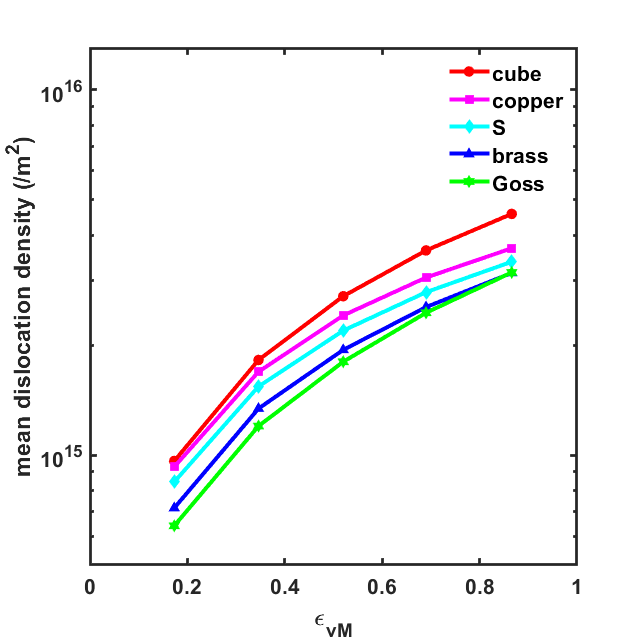} 
    \label{dd_evolution}
}
\subfigure[]{
 \includegraphics[height=0.5\linewidth]{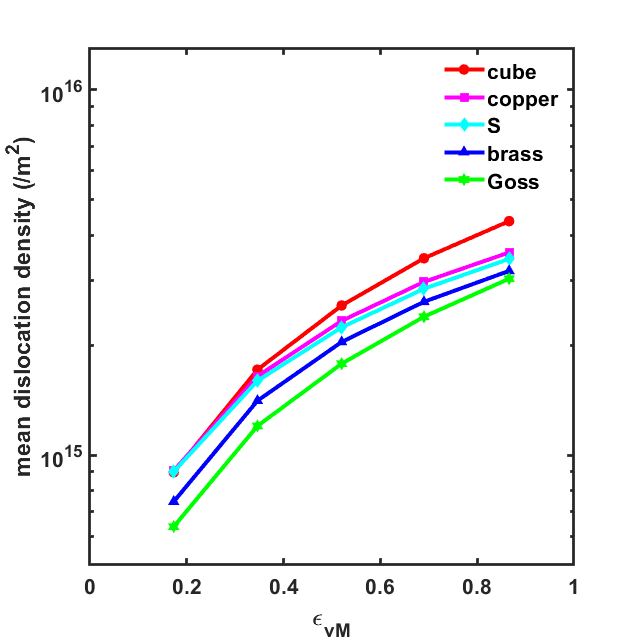}
}
\caption{Evolution of the mean dislocation density with increasing strain for the main texture components in (a) RVE1 and (b) RVE2. For both the plots logarithmic scale is used in Y-axis.}
\label{}
\end{figure}

In Figure \ref{dd_evolution} the mean dislocation density of the main texture components are plotted with increasing strain. In both RVE1 and RVE2, on an average the cube oriented regions accumulate more dislocation density than any other texture component. The copper component has the second highest dislocation density and the the difference between dislocation density in the cube and copper component
increases with increasing strain. Moreover, the brass and Goss components have the lowest average dislocation density. This result corroborates well with the X-ray difraction findings of \cite{szpunar,kallend}. Additionally, \cite{godfrey} used transmission electron microscopy (TEM) to estimate the stored energy in copper, S and brass oriented regions in cold rolled aluminum.  They found that stored energy is ranked with copper $>$ S $>$ brass. 

Although Kallend and Huang\cite{kallend} made an attempt to connect Taylor factor of different texture components with their stored energy, Taylor factor may not be a good indicator of stored energy. Our simulation results show that cube has a higher dislocation density (equivalently stored energy) than other texture components despite having a lower Taylor factor. In reality, the cube oriented grains are not stable and non-cube grains, which may not have low Taylor factor, can rotate towards cube orientation during deformation. Apart from that, Huang and Kallend were unable to measure the stored energy of cube component with high accuracy due to its low volume fraction. In a later work, using the same method proposed by Huang and Kallend, Rajmohan and Szpunar \cite{szpunar} showed that cube component indeed has high stored energy. Therefore, Taylor factor analysis, which assumes iso-strain condition, may not provide an accurate estimate for the stored energy of the cube oriented regions which are located near large strain heterogeneities such as grain boundaries (see Figure \ref{TB_band}). 

\begin{figure}[!htbp]
    \centering
    \includegraphics[height=0.55\linewidth]{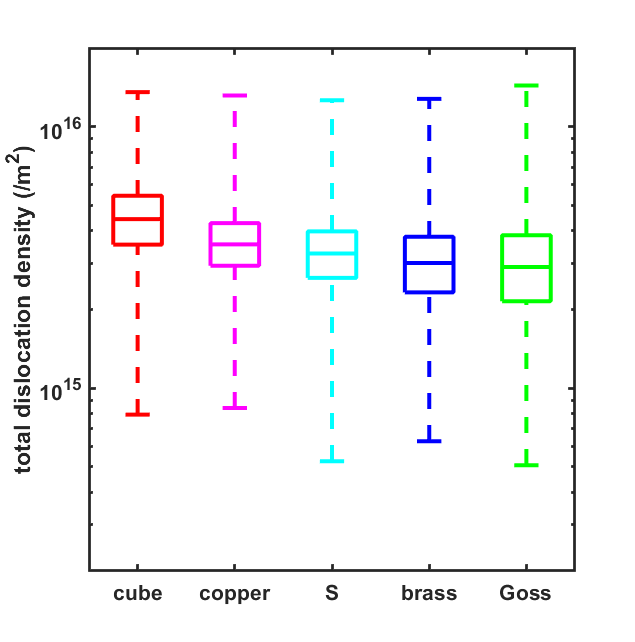}
\caption{Dislocation density distribution for the main texture components at true strain of $\epsilon_{vM}$ = 0.87 are presented as box plots. The lower, middle and upper limits of the box plot represents the first, second and third quartile of the distribution, respectively. Range of the distribution is also shown as error bars on the box plots. Logarithmic scale is used for Y-axis.}
\label{dd_spread}
\end{figure}

Although it may be apparent that cube component will have a nucleation advantage over other components due to its high average stored energy, to formulate a strong conclusion we need to look into the actual distribution of dislocation density in each of the texture components. In Figure \ref{dd_spread} the dislocation density distribution for the main texture components at macroscopic strain of $\epsilon_{vM} = 0.87$ are presented as box plots. The lower, middle and upper limits of the box plot represents the first, second (median) and third quartile of the distribution, respectively. The range of the dislocation density (minimum to maximum value) is also shown as an error bar on each of the box plot. The figure shows that dislocation density in all the components is widely spread with maximum values in the same range. Among all the main texture components, the Goss component has the largest spread. Even though cube component has the highest median dislocation density, it also has low volume fraction in the deformed state. Therefore, it is expected that nucleation will occur from all the texture components during static recrystallization.

\begin{figure}[!htbp]
    \centering
    \includegraphics[height=0.5\linewidth]{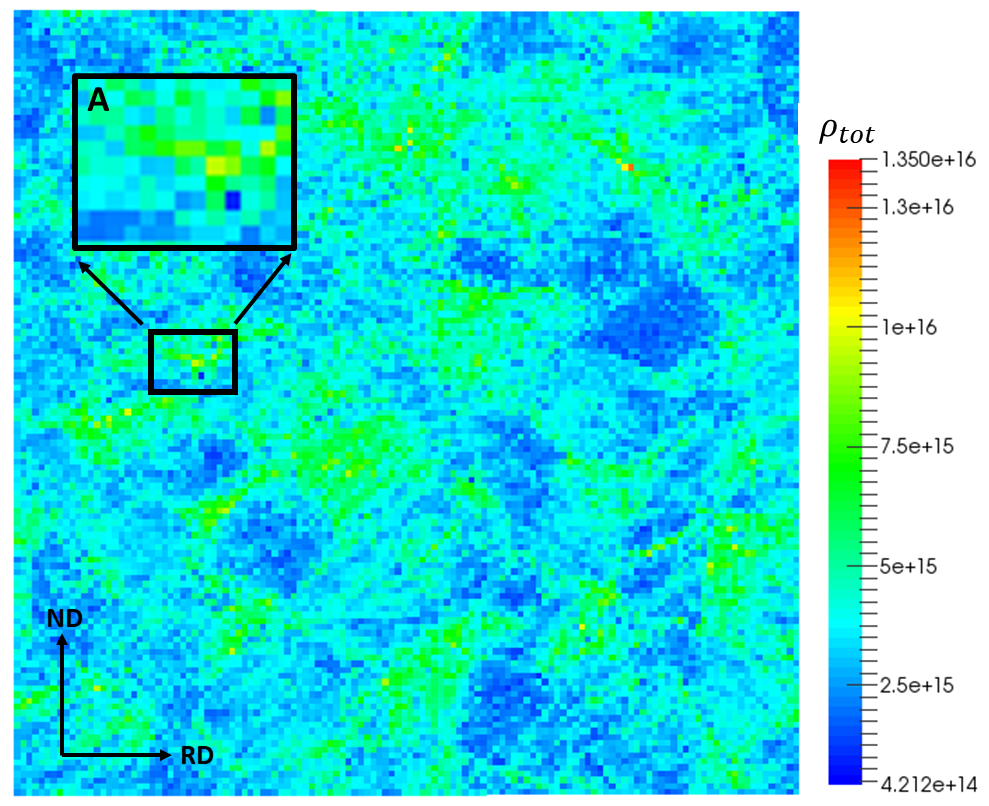}
    \caption{Dislocation density distribution at true strain of $\epsilon_{vM}$ = 0.87 is shown in a 2D section. The colorbar shows the magnitude of dislocation density in units of ($/m^2$). The region shown in Figure \ref{TB_band} is highlighted here with a black box. Inset \lq A' shows the same region with higher magnification.}
    \label{dislocation_spatial}
\end{figure}

The spatial distribution of dislocation density is highly heterogeneous in the deformed microstructure. In Figure \ref{dislocation_spatial}, total dislocation density at true strain of $\epsilon_{vM}$ = 0.87 is plotted in a 2D section along the transverse axis. Dislocation density variation of two orders of magnitude can be observed in this figure. Bands of high dislocation density (hotspots) align mainly along the RD or along a certain angle with RD. On the other hand, patches of low dislocation density (coldspots) can be observed in some of the grain interior regions. The cube band region shown in Figure \ref{TB_band} is also highlighted with a black box. The same region is magnified in inset \lq A' which shows a high accumulated dislocation density. As dislocation density (in turn, stored energy) controls both nucleation and grain growth during recrystallization, the heterogeneous distribution of dislocation density can have profound influence on recrystallization kinetics, which will be discussed later. 

\subsection{Static recrystallization simulation}
\label{rex_static}
As we discussed earlier, the static recrystallization process includes concurrent nucleation and growth of recrystallized grains. In this work, we performed static recrystallization simulations after 50\% rolling reduction ($\epsilon_{vM}$ = 0.87) at high annealing temperature (grain boundary mobility at 450\degree C \cite{vandermeer} is used). Effect of various rolling reductions and annealing temperatures will be addressed in a future work. 

In an earlier work by Mohamed and Bacroix \cite{bacroix} it was shown that for a given stored energy a critical annealing temperature is required below which the material will not recrystallize. Nucleation by accelerated recovery processes is less likely to occur below this critical temperature. As the rate of dislocation climb and cross slip controlled recovery processes increase with increasing temperature, probability of accelerated recovery in the high stored energy locations increases with increasing temperature. Therefore, values for Weibull scale parameter ($k_c$), shape parameter ($\eta$) and nucleation attempt frequency ($u^{nuc}$) should be chosen in such a manner that nucleation rate would increase with increasing temperature. Here, the values for $k_c$, $\eta$ and $u^{nuc}$ are selected in such a way that all the nuclei get selected from the top 25\% (above third quartile) of the dislocation density distribution. The nucleation model parameters are given in Table \ref{nucleation_para}.
\begin{table}[!htbp]

\caption{Parameters used for nucleation  model.}
\label{nucleation_para}
 \begin{tabular}{ll}
 \hline
 parameter & value\\
 \hline
    $k_c$ & $9.0\times10^{15}\quad m^{-2}$\\
    $\eta$ & 13.0 \\
    $u^{nuc}$ & $4 s^{-1}$\\
\hline
\end{tabular} 
\end{table}

Other than nucleation rate, grain boundary mobility also increases with increasing temperature. Hence, we adopted the temperature dependent mobility parameters from the work of Vandermeer et al. \cite{vandermeer}. To simulate recrystallization grain growth we used isotropic grain boundary energy and mobility in our phase field model. Although the deformed grains developed significant orientation gradients, for the present work we still represent each of the deformed grains with a single order parameter.  The phase field model parameters used in this work are listed in Table \ref{pf_para}.
\begin{table}[!htbp]

\caption{Parameters used for phase field simulation.}
\label{pf_para}
 \begin{tabular}{ll}
 \hline
    parameter & value \\
\hline
     $m$ & $4.0\times10^5\quad Jm^{-3}$ \\
     $\Gamma$ & 1.0 \\
     $kappa$ & $1.5\times10^{-6}\quad Jm^{-1}$\\
     $E_{gb}$ & $0.6 \quad Jm^{-2}$ \\
     $M_{gb}$ & $1.2\times10^{-11}\quad m^4J^{-1}s^{-1}$ \cite{vandermeer} \\
     $\Delta x$ & $3.0\times10^{-6}\quad m$ \\
     $\Delta t^{pf}$ & $0.004 s$\\
     $\delta$ & 3\\
\hline
\end{tabular} 
\end{table}

\subsubsection{Nucleation}
\label{rex_nucleation}
\begin{figure}[!htbp]
    \centering
\subfigure[]{
    \includegraphics[height=0.5\linewidth]{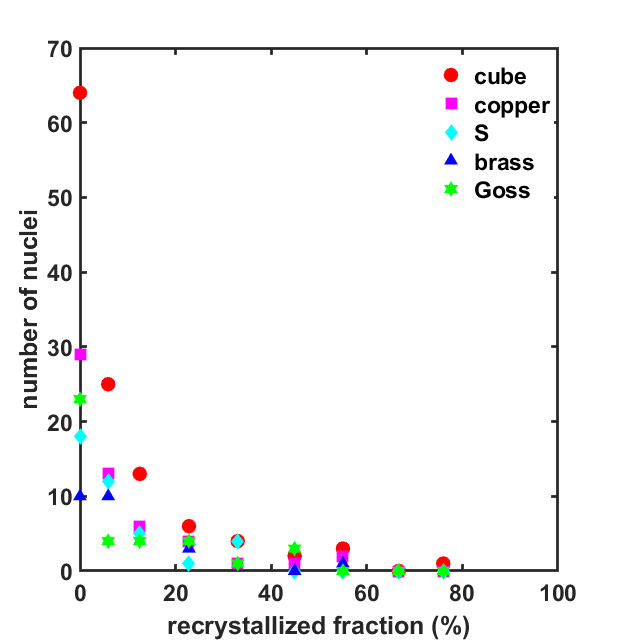}
  
    \label{nuc_RVE1}
}
\subfigure[]{
    \centering
    \includegraphics[height=0.5\linewidth]{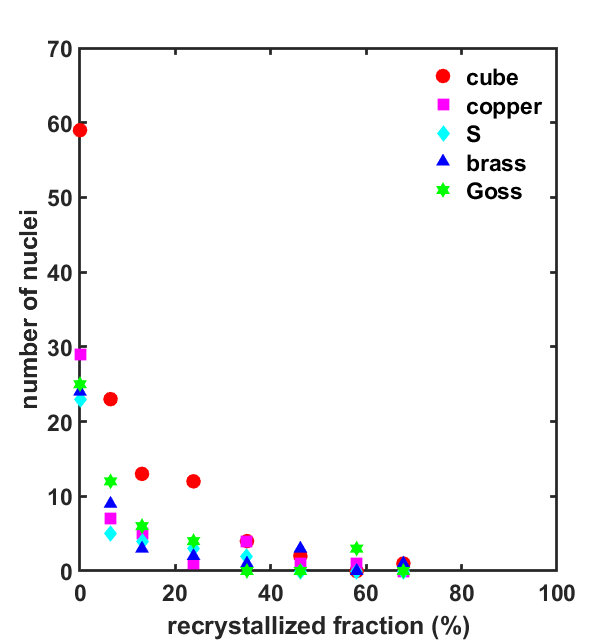}
   
    \label{nuc_RVE2}
}
\caption{Frequency of nucleation for the main texture components at various stages of recrystallization in (a) RVE1 and (b) RVE2.}
\label{nucleation_sequence}
\end{figure}

 Figure \ref{nucleation_sequence} shows the nucleation number frequency for the main texture components with increasing recrystallization volume fraction. It is clear that for both RVE1 and RVE2 nucleation frequency is high during the initial stages of recrystallization and rapidly decreases with the progress of recrystallization. Essentially, the final 20\% of volume transformation occurs without any new nucleation. Among all the main texture components, cube component has the highest frequency for both RVE1 and RVE2 due to predominant nucleation during the initial stages of recrystallization. Copper and Goss oriented nuclei are also present in significant numbers. 

In RVE1 a total 1109 grains nucleated, of which only 118 grains had a cube orientation. For RVE2 a total 1196 grains nucleated of which only 115 grains had a cube orientation. To verify that these are significant numbers relative to the other texture components, we follow the approach of Samajdar et al. \cite{samajdar1999} and Samajdar and Doherty \cite{samajdar1998,doherty1985} to analyze the nucleation propensity of various texture components. We define $A_i$ as the nucleation frequency of a particular texture component $i$ and $A_r$ as the nucleation frequency expected for any texture component when nucleation is completely random. According to \cite{doherty1997current,doherty1985}, for preferential nucleation to occur the ratio of $\tfrac{A_i}{A_r}$ needs to be much higher than 1. For any texture component measured withing 
$15\degree$ from the ideal orientation, random nucleation of that component corresponds to a nucleation frequency of 2.5\% of the total number of nuclei \cite{doherty1997,mackenzie}. Therefore, for any texture component in RVE1 $A_r$ would be $1109\times 0.025 = 27.7$ and in RVE2 $A_r$ would be $1196\times 0.025 = 29.9$. Values of $A_i$ and $\tfrac{A_i}{A_r}$ for both RVE1 and RVE2 are presented in Table \ref{nucleation_propensity}. The cube component has $\tfrac{A_i}{A_r}$ ratio of 4.2 for RVE1 and 3.8 for RVE2, which is fairly higher than 1.0. Additionally, among all the main texture components, the cube component has the highest $\tfrac{A_i}{A_r}$ ratio.  Therefore, we can infer that the cube component has a nucleation advantage over other texture components. Interestingly, despite having the lowest mean dislocation density the Goss component has higher propensity of nucleation than the S component in both RVE1 and RVE2. Due to large spread in the dislocation density distribution (see Figure \ref{dd_spread}) some Goss oriented regions possess significantly high dislocation density which increase the propensity of Goss oriented nucleation.  

\begin{table}[!htbp]

\caption{Analysis of nucleation frequency advantage.}
\label{nucleation_propensity}
 \begin{tabular}{lllll}
 \hline
\multirow{3}{*}{component} & \multicolumn{2}{c}{RVE1} & \multicolumn{2}{c}{RVE2}\\
 \cline{2-5} &
   \multicolumn{1}{c}{$A_i$} &
   \multicolumn{1}{c}{$\dfrac{A_i}{A_r}$}  &
   \multicolumn{1}{c}{$A_i$} & \multicolumn{1}{c}{$\dfrac{A_i}{A_r}$}  \\
\hline
     cube & 118 & 4.2 & 115 & 3.8 \\

     copper & 56 & 2.0 & 48 & 1.6 \\

     S & 40 & 1.4 & 37 & 1.2\\

     brass & 29 & 1.0 & 43 & 1.4\\
    
     Goss & 39 & 1.4 & 50 & 1.7\\
\hline
\end{tabular} 

\end{table}

\begin{figure}[!htbp]
    \centering
    \subfigure[$\epsilon_{vM}$ = 0.0]{
    \includegraphics[height=0.45\linewidth]{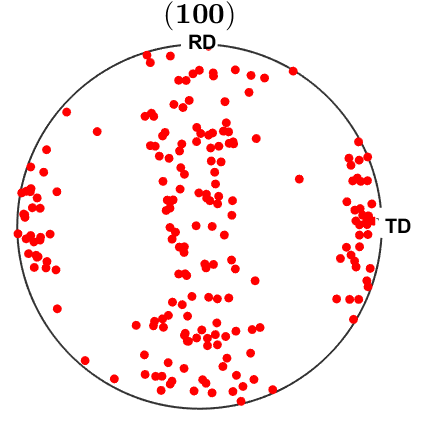}
    }
    \subfigure[$\epsilon_{vM}$ = 0.52]{
    \includegraphics[height=0.45\linewidth]{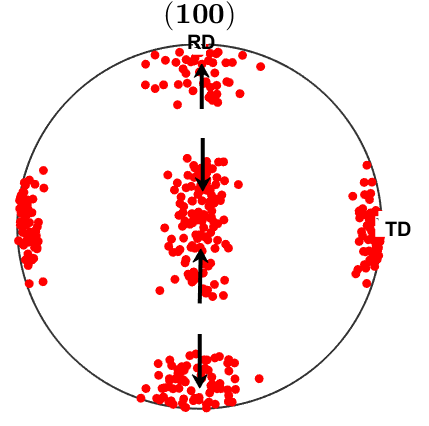}
    }
    \subfigure[$\epsilon_{vM}$ = 0.87]{
    \includegraphics[height=0.45\linewidth]{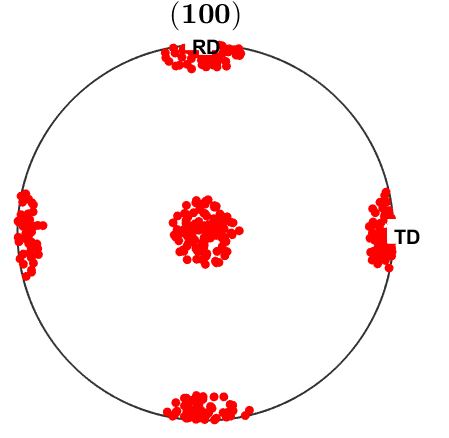}
    }
    \caption{\{100\} pole figures showing the evolution of crystallographic orientation at locations where cube nuclei have been found. Each marker represents a single orientation. In the undeformed state (a), transverse direction was already aligned to the cube direction. With increasing strain the orientations rotate around TD (b) and finally reach the ideal cube orientation (c). Rotation directions are marked with arrows in (b).}
    \label{rotation}
\end{figure}

In section \ref{deformation}, we discussed that cube grains are not stable in plane strain compression and most of the retained cube fraction comes from initially non-cube orientations. Here we study the crystallographic origin of the cube nuclei in more detail. In Figure \ref{rotation} we tracked the lattice rotation paths for all the cube oriented nuclei in  RVE1. In the \{100\} pole figures, each marker represents a crystallographic orientation corresponding to each location where cube nucleation occured. One can observe that initially, in most of the locations, the transverse direction (TD) was already aligned with cube direction. On the other hand, only in few locations the normal direction (ND) and rolling direction (RD) were aligned to the cube direction. Therefore, during deformation, majority of these orientations rotate around TD in order to gradually align ND and RD planes with the cube direction. In case of RVE2, a similar lattice rotation around TD is observed for the cube nuclei.

These findings are in contrast with Dillamore and Katoh mechanism \cite{Dillamore} in which it was hypothesized that cube oriented recrystallized grains would nucleate from transition bands which are stable around ND rotation. Although fragments of ND cube orientation can be found in the microstructure, they are not necessarily efficient for nucleation. Additionally, all the cube nuclei were located near the grain boundary region as shown in Figure \ref{TB_band}. Therefore, nucleation of cube grains were preferred in the intergranular regions where strain heterogeneities are maximum. A similar type of intergranular cube band nucleation has been reported by Albou et al. \cite{albou2010} in Al-Mn alloy where they showed intergranular nucleation is more efficient than transgranular cube band nucleation. A separate study is necessary to fully understand which stress state, slip activity and grain orientation favors the formation of transgranular and intergranular cube bands. 

Another popular hypothesis for the preferential nucleation of cube grains, is that cube oriented subgrains recover rapidly due to the presence of orthogonal dislocation pairs \cite{Ridha}. However, the hypothesis that orthogonal dislocation pairs will help only cube subgrains to preferentially nucleate has not been rigorously tested. Therefore, we analyzed the slip system activity in all locations where nucleation has taken place and counted the number of orthogonal slip system pairs for each of those locations. In fcc metals, 9 orthogonal dislocation pairs are possible for 12 independent slip systems. In this analysis, for a macroscopic strain of $\epsilon_{vM} = 0.87$, an orthogonal slip system pair has been counted if the accumulated shear strain was greater than 0.2 for both the slip systems. For example, [1 0 1](1 -1 -1) and [1 0 -1](1 1 1) will be counted as an active orthogonal slip system pair if both of them have shear strain of 0.2 or more.

\begin{figure}[!htbp]
    \centering
    \subfigure[]{
    \includegraphics[height=0.5\linewidth]{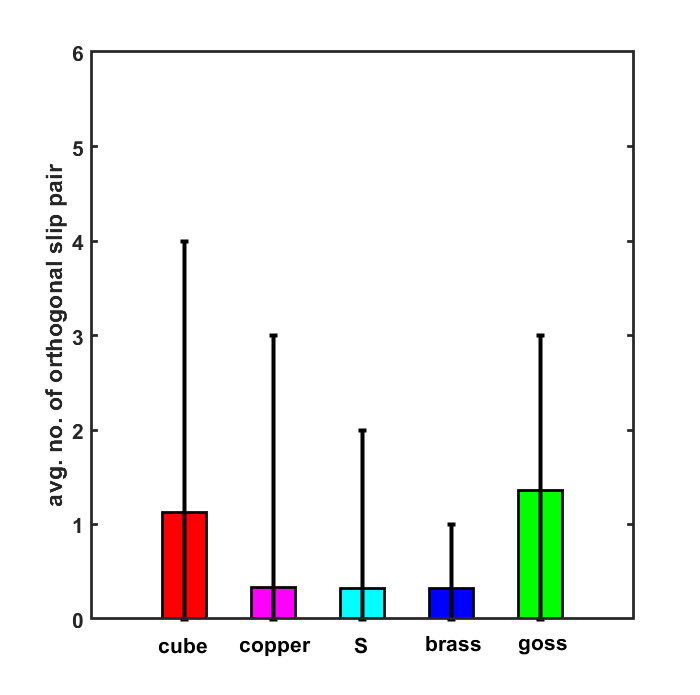}
    
    \label{fig:my_label}
    }
\subfigure[]{
    \includegraphics[height=0.5\linewidth]{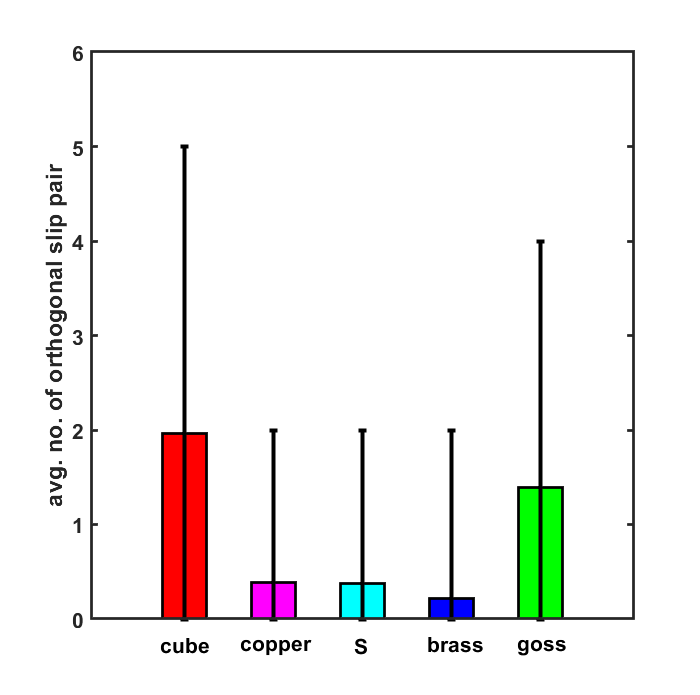}
    
    \label{avg_ortho}
    }
    \caption{Bar plots corresponding to main texture components showing the average number of orthogonal slip system pairs active at the nuclei locations in (a) RVE1 and (b) RVE2. Error bars show the range of the distribution.}
    \label{fig:my_label}
\end{figure}

In Figure \ref{avg_ortho} we plot the average number of orthogonal slip pairs found at the locations where nucleation has occurred. The range of the distribution (minimum to maximum value) is also shown as error bar. For both RVE1 and RVE2, cube oriented nuclei have the highest number of orthogonal slip pairs, whereas copper, S and brass oriented nuclei have the lowest number of orthogonal slip pairs. Perhaps the most surprising finding of this analysis is that number of orthogonal slip system pair in the Goss oriented nuclei is comparable to that in cube oriented nuclei. Hence, orthogonal slip activity is not unique to the cube component as previously been thought of (\cite{Ridha}), rather both cube and Goss component would nucleate preferentially if orthogonal slip activity helps in faster recovery. Therefore, we can infer that higher stored energy in the cube component is the most important factor for the preferential nucleation of cube grains.     

\subsubsection{Grain growth}
\label{rex_growth}
\begin{figure}[!htbp]
    \centering
    \subfigure[]{
    \includegraphics[height=0.5\linewidth]{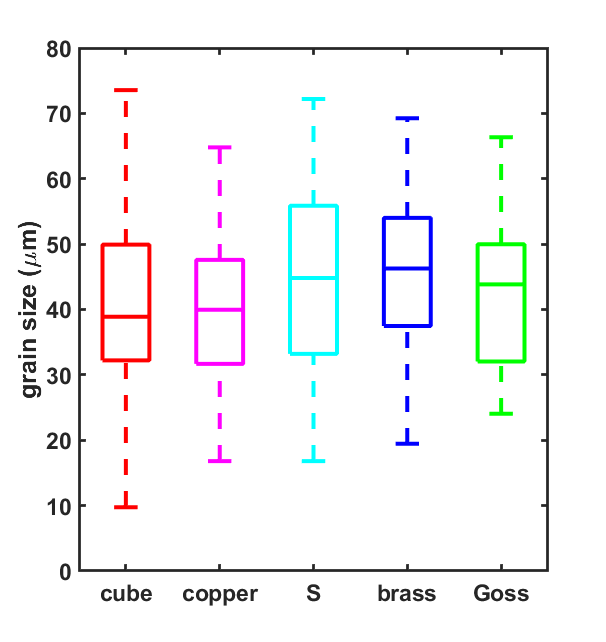}
    \label{fig:my_label}
    }
     \subfigure[]{
    \includegraphics[height=0.5\linewidth]{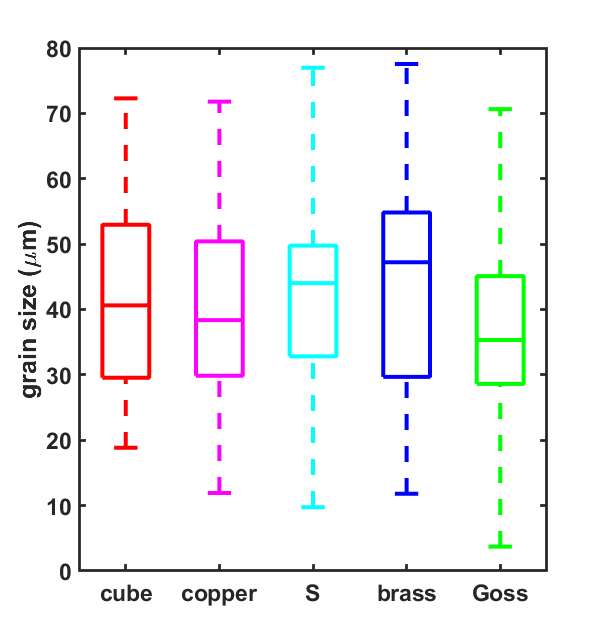}
    \label{fig:my_label}
    }
 \caption{Grain size distribution for the main texture components in (a) RVE1 and (b) RVE2 are presented as box plots. The lower, middle and upper limits of the box plot represents the first, second and third quartile of the distribution, respectively. Range of the distribution is also shown as error bars on the box plots.}
    \label{grain_size}   
\end{figure}

In Figure \ref{grain_size} the grain size distribution after full recrystallization is presented as a box plot for the main texture components. Error bars in the figure show the range of the grain size distribution. Grain size is reported as the equivalent spherical diameter (ESD) calculated from the volume of each grain. In both RVE1 and RVE2, the grain size distribution is widely spread for all the components. Apart from the brass component which has slightly higher median grain size, other texture components have more or less the same median grain size. The median grain size in RVE1 is 42.6 $\mu$m and in RVE2 is 41.1 $\mu$m. Hence, the median grain size of all the texture components are similar to the median grain size of the whole microstructure. Additionally, the maximum grain size for all the texture components are in the same range for RVE1, whereas the Goss component has slightly larger maximum grain size in RVE2. Therefore, we can infer that there is no preferential growth for the cube or any other texture component in our study.      

In a previous work, Lauridsen et al. \cite{lauridsen2003} used 3DXRD technique to study recrystallization kinetics in cold rolled aluminum in which they reported grain size distribution, mean grain size and grain growth rate for cube, rolling and other texture components. Their results show that cube and rolling texture components have similar grain size distribution and mean grain size which is also the case in our study. Although they concluded that cube component has a growth advantage, we believe that large cube grains can be observed due to lack of impingement from other grains. Nes and coworkers \cite{hjelen1991,solberg} reported heterogeneous nucleation throughout the deformed matrix and presence of dense nuclei clusters during initial annealing period. Clustering of nuclei was also reported in the simulation works of \cite{vandermeer1995,jensen1997,marthinsen,raabe2000}. 
\begin{figure}[!htbp]
    \centering
    \subfigure[22.8\% recrystallized]{
    \includegraphics[height=0.45\linewidth]{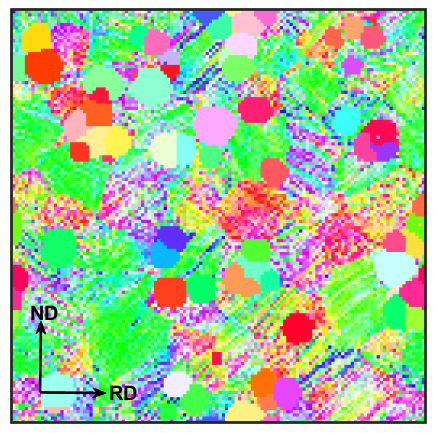}
    \label{fig:my_label}
    }
     \subfigure[44.9\% recrystallized]{
    \includegraphics[height=0.45\linewidth]{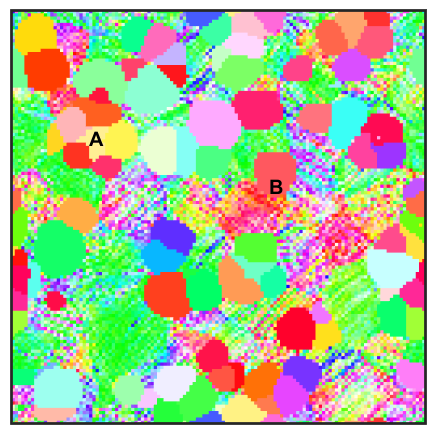}
    \label{nuclei_cluster}
    }
    \subfigure[83.4\% recrystallized]{
    \includegraphics[height=0.45\linewidth]{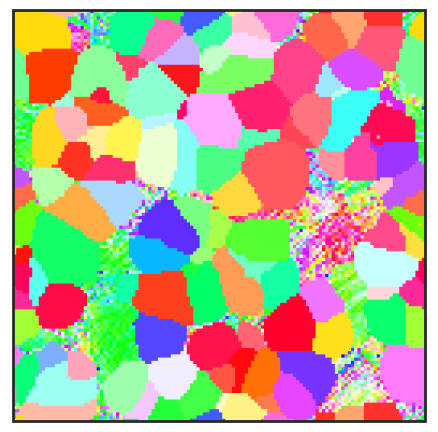}
    \label{fig:my_label}
    }
   \subfigure[100\% recrystallized]{
    \includegraphics[height=0.45\linewidth]{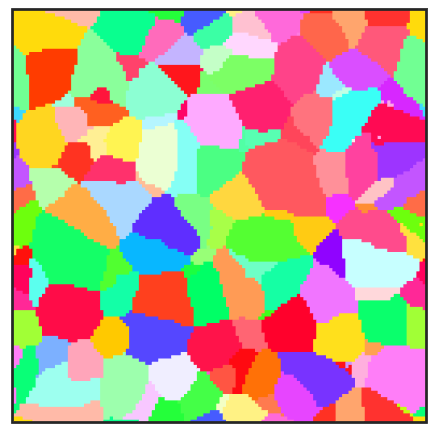}
    \label{fig:my_label}
    }
   \caption{Microstructure evolution during recrystallization is shown in a 2D slice from RVE1. (a), (b), (c) and (d) correspond to the same slice at various stages of recrystallization. In (b) a cluster of nuclei can be observed near the location marked as A and an isolated cube grain can be observed near the location B. [0 0 1] IPF color code is used to represent the orientations.}
    \label{microstructure_evolution} 
\end{figure}

In Figure \ref{microstructure_evolution} we present microstructure evolution during recrystallization in a 2D slice extracted from the 3D microstructure RVE1. Two different locations of interest have been highlighted with letters \lq A' and \lq B' in Figure \ref{nuclei_cluster}. A cluster of recrystallized nuclei can be observed at location \lq A', whereas at location \lq B' an isolated cube oriented (within 15\degree from ideal orientation) recrystallized grain can be spotted. Due to clustering all the nuclei around \lq A' quickly impinge on each other, resulting in hindered grain growth at that location. Conversely, due to lack of nucleation around \lq B' the cube oriented grain grow without any obstruction for a long period of time. Therefore, spatial distribution of the nuclei is an important factor for the final grain size distribution. Although here we showed growth of large cube grains, non-cube grains can also grow large due to lack of impingement. This is why Figure \ref{grain_size} shows similar grain size distribution for both the cube and other texture components.

In a recent study, Lin et al. \cite{lin} reported growth of large cube grains during static annealing of cold rolled aluminum. Although they argued that these cube grains had a growth advantage due to a special misorientation (45\degree-50\degree and $\langle1 1 1\rangle$ axis) with the deformed matrix, other cube grains with smaller grain size had a similar misorientation distribution. Therefore, the effect of misorientation on growth rate is not clear from their study. On the other hand, the effect of impingement (or lack of impingement) can not be ignored because the cube grains which grew large must be growing without impingement from other grains. Nevertheless, in a future work misorientation dependent grain boundary energy and mobility can be implemented in the phase field model (\cite{kazaryan1,kazaryan2}) to study their effects on recrystallization grain growth.

\subsubsection{Recrystallization texture evolution}
\label{rex_tex_evolution}
\begin{figure}[!htbp]
    \centering
    \subfigure[]{
    \includegraphics[height=0.5\linewidth]{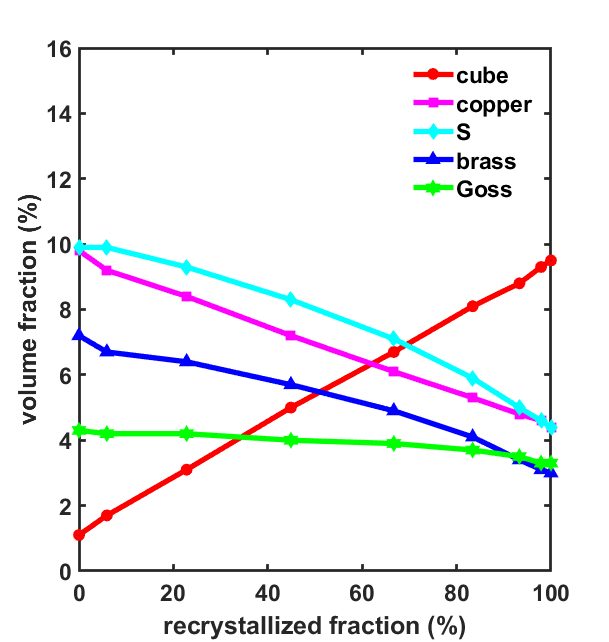}
    \label{fig:my_label}
    }
\subfigure[]{
    \includegraphics[height=0.5\linewidth]{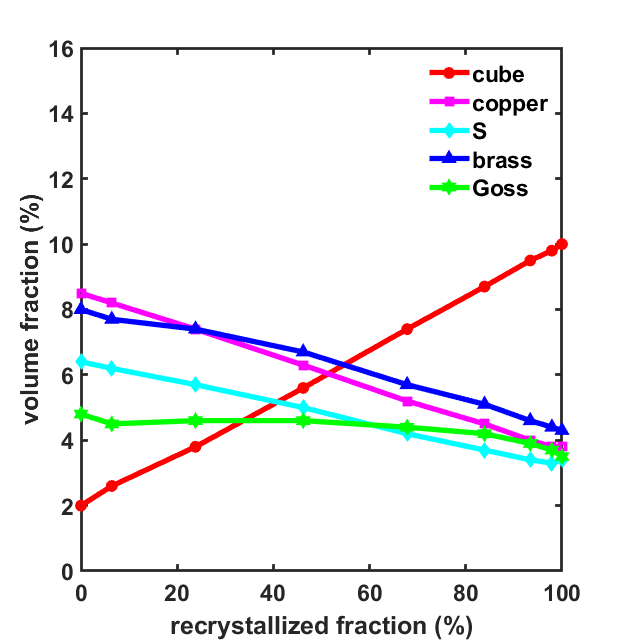}
    \label{fig:my_label}
    }
    \caption{Evolution of the main texture components with increasing recrystallized fraction for (a) RVE1 and (b) RVE2.}
    \label{rex_texture}
\end{figure}

As the recrystallized grains grow in the deformed matrix, the recrystallization texture starts to replace the deformation texture. Figure \ref{rex_texture} shows the evolution of the main texture components during recrystallization. While the volume fraction of the cube component increases with increasing recrystallization fraction, the volume fraction of the rolling texture components (copper, S, brass) decreases. Volume fraction of the Goss component remains more or less constant during the recrystalization process.

\begin{figure}[!htbp]
    \centering
\subfigure[]{
    \includegraphics[height=0.5\linewidth]{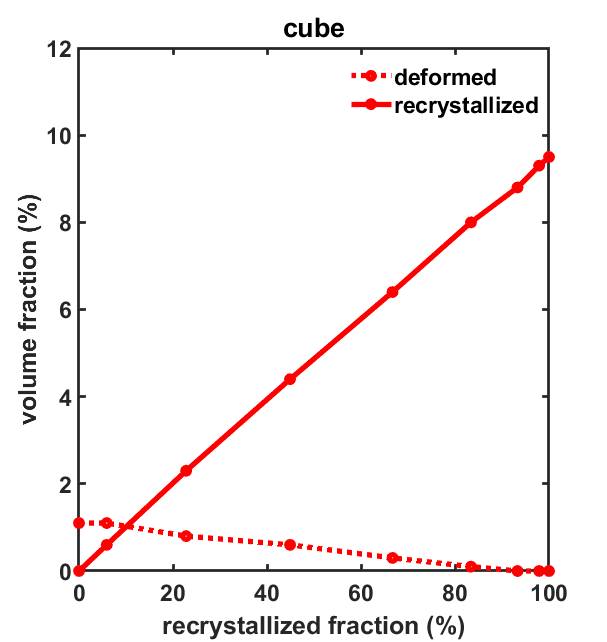}
    \label{fig:my_label}
    }
\subfigure[]{
    \includegraphics[height=0.5\linewidth]{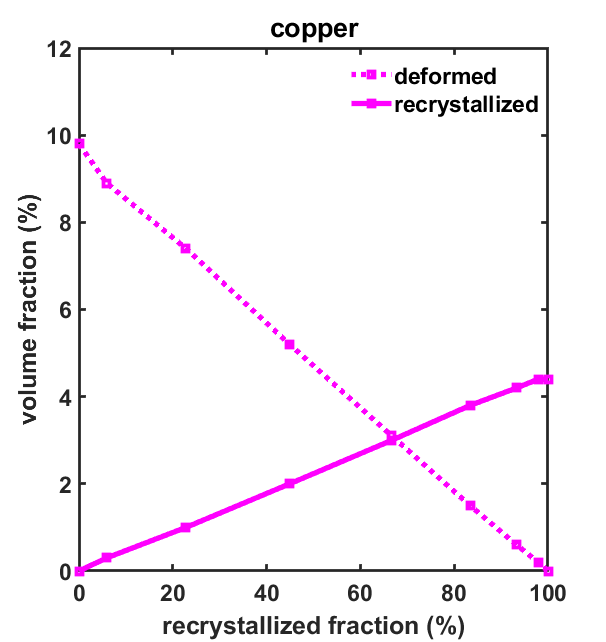}
    \label{fig:my_label}
    }
\subfigure[]{
    \includegraphics[height=0.5\linewidth]{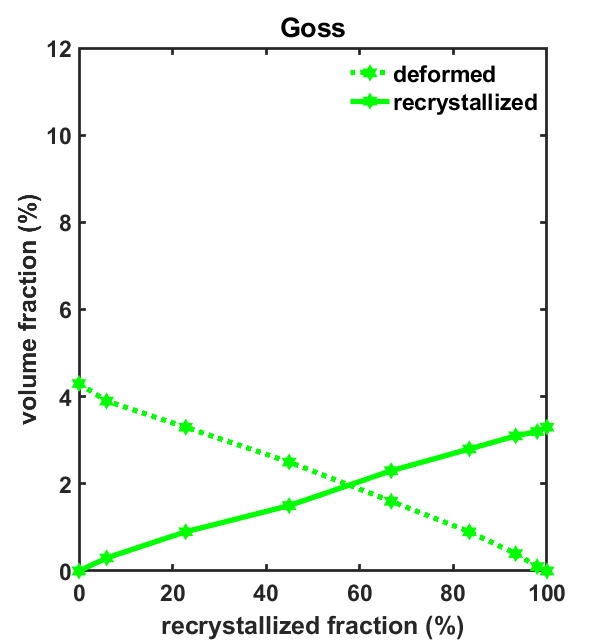}
    \label{fig:my_label}
    } 
    \caption{Volume fraction changes in the deformed and recrystallized portions of (a) cube, (b) copper and (c) Goss components during recrystallization process.}
    \label{deformed_rex}
\end{figure}

These changes in volume fraction primarily depend on two competing factors : i) volume fraction in the deformed state which decreases during the recrystallization process and ii) volume fraction in the recrystallized grains which increases during the recrystallization process. In Figure \ref{deformed_rex} we plot the evolution of the deformed and recrystallized portions of the cube, copper and Goss component. The cube component volume fraction in the deformed state was initially low. However during recrystallization, the volume fraction in the recrystallized portion increases rapidly mainly due to preferred nucleation of cube oriented grains. Therefore, the volume fraction decrement during recrystallization (due to deformed cube regions being consumed) is negligible in comparison to the volume fraction increment. Hence, a steady increase in the overall volume fraction can be observed for the cube component. 

For the copper component the decrease in the deformed material is faster than the increase in recrystallized portion. Due to high average stored energy in the deformed state (Figure \ref{dd_evolution}) the copper oriented regions are quickly invaded by the recrystallized grains. On the other hand, the rate of increase in the recrystallized fraction is low due to inadequate nucleation. Therefore, a steady decrease in volume fraction is observed for the copper component. Behavior of the other rolling texture components such as S and brass components are also similar to the copper component. However, the brass component has a slightly lower rate of decrease (see Figure \ref{rex_texture}). This is probably due to its lower stored energy in the deformed state relative to copper and S.

Interestingly, for the Goss component the rate of decrease in the deformed fraction is comparable to the rate of increase in the recrystallized fraction. One possible reason for this behavior might be the low average stored energy in the Goss component (see Figure \ref{dd_evolution}). Low stored energy facilitates slower removal of the deformed volume fraction. Moreover, large spread in the stored energy distribution (see Figure \ref{dd_spread}) results in significant nucleation of Goss oriented grains; which in turn helps to increase the recrystallized volume fraction. Thus, volume fraction of the Goss component remains more or less constant during the whole recrystallization period. 

\section{Conclusions}
The intention of this work was to understand the origin of cube texture development during static recrystallization of cold rolled fcc metals. We used an EVP-FFT based crystal plasticity model with dislocation density based constitutive laws to simulate plane strain compression of copper. A stochastic nucleation model and phase field grain growth model were used to simulate static recrystallization after 50\% rolling reduction. Main findings of this study can be summarized as following:

\begin{enumerate}

\item During plane strain compression, a typical \emph{copper type} rolling texture evolved in which volume fraction of the copper, S and brass texture components increased and volume fraction of the cube component decreased. However, if the initial microstructure contains a relatively small cube component then a small increase can be observed during intermediate stages of deformation. This increase strongly suggests that grains which were initially non-cube oriented can rotate towards cube during deformation. The cube oriented regions originating from non-cube grains were generally found as narrow intergranular bands surrounded by rolling and random (other than rolling) texture components. 

\item Cube regions had highest dislocation density followed by copper, S, brass and Goss. This trend in dislocation density is in good agreement with previous experiments. These observed dislocation densities are not consistent with the iso-strain Taylor factor analysis for plane strain compression.

\item Initial cube volume fraction had little effect on the nucleation of cube orientations. Instead, almost all of the cube nuclei originated from initially non-cube grains. 

\item These non-cube grains mainly rotated around TD to reach the ideal cube orientation. Dillamore and Katoh \cite{Dillamore} theorized that cube nuclei would originate from transition bands which are stable around ND rotation. Therefore, our simulation results are not in agreement with the proposed Dillamore and Katoh mechanism. 

\item Orthogonal dislocation pairs were found in both cube and Goss components. Therefore, our simulation results do not support the hypothesis that only cube subgrains can preferentially recover during annealing.   

\item Cube component had clear nucleation advantage over the other texture components. In contrast, none of the texture component showed any growth advantage. The average grain size was more or less same for all the texture components. 

\item Instead of a growth advantage, the presence of large cube grains can be attributed to heterogeneous nucleation. Nuclei which formed in clusters impinged rapidly on each other and were not able to grow. On the other hand, isolated nuclei grew large due to lack of impingement from other nuclei.

\item Significant increase in cube volume fraction was observed during recrystallization due to low volume fraction of cube component in deformed state as well as high cube nucleation frequency.
\end{enumerate}

\section*{Acknowledgements}
The authors want to acknowledge the National Science Foundation, Division of Civil, Mechanical and Manufacturing Innovation for their support under the Grant No. CMMI-1662646. SC and CSP want to acknowledge the Simulation Innovation and Modeling Center, The Ohio State University and Ohio Supercomputer Center for providing computational resources. SC and CSP want to thank Dr. Pengyang Zhao for providing his dynamic recrystallization code. SC and CSP also want to thank Prof. Anthony Rollett for encouraging them to investigate cube texture development.  

\bibliography{mybibfile}

\begin{thebibliography}{10}
\expandafter\ifx\csname url\endcsname\relax
  \def\url#1{\texttt{#1}}\fi
\expandafter\ifx\csname urlprefix\endcsname\relax\def\urlprefix{URL }\fi
\expandafter\ifx\csname href\endcsname\relax
  \def\href#1#2{#2} \def\path#1{#1}\fi

\bibitem{kocks1998}
U.~F. Kocks, C.~N. Tom{\'e}, H.-R. Wenk, Texture and anisotropy: preferred
  orientations in polycrystals and their effect on materials properties,
  Cambridge university press, 1998.

\bibitem{suwas2014}
S.~Suwas, R.~K. Ray, Crystallographic texture of materials, Springer, 2014.

\bibitem{tucker1961}
G.~Tucker, Texture and earing in deep drawing of aluminium, Acta Metallurgica
  9~(4) (1961) 275--286.

\bibitem{engler1996}
O.~Engler, J.~Hirsch, Recrystallization textures and plastic anisotropy in
  al-mg-si sheet alloys, in: Materials Science Forum, Vol. 217, Trans Tech
  Publ, 1996, pp. 479--486.

\bibitem{matsuo1989}
M.~Matsuo, Texture control in the production of grain oriented silicon steels,
  ISIJ international 29~(10) (1989) 809--827.

\bibitem{dorner2007}
D.~Dorner, S.~Zaefferer, D.~Raabe, Retention of the goss orientation between
  microbands during cold rolling of an fe3\% si single crystal, Acta materialia
  55~(7) (2007) 2519--2530.

\bibitem{bhattacharjee2007nickel}
P.~P. Bhattacharjee, R.~K. Ray, A.~Upadhyaya, Nickel base substrate tapes for
  coated superconductor applications, Journal of materials science 42~(6)
  (2007) 1984--2001.

\bibitem{merlini1953}
A.~Merlini, P.~Beck, Study of the origin of the cube texture, Acta Metallurgica
  1~(5) (1953) 598--606.

\bibitem{Ridha}
A.~Ridha, W.~Hutchinson, Recrystallisation mechanisms and the origin of cube
  texture in copper, Acta metallurgica 30~(10) (1982) 1929--1939.

\bibitem{hjelen1991}
J.~Hjelen, R.~{\O}rsund, E.~Nes, On the origin of recrystallization textures in
  aluminium, Acta metallurgica et materialia 39~(7) (1991) 1377--1404.

\bibitem{duggan1993}
B.~Duggan, K.~L{\"u}cke, G.~K{\"o}hlhoff, C.~Lee, On the origin of cube texture
  in copper, Acta metallurgica et materialia 41~(6) (1993) 1921--1927.

\bibitem{humphreys2012}
F.~J. Humphreys, M.~Hatherly, Recrystallization and related annealing
  phenomena, Elsevier, 2012.

\bibitem{necker1970}
C.~Necker, R.~Doherty, A.~Rollett, Quantitative measurement of the development
  of recrystallization texture in ofe copper, Textures and Microstructures 14
  (1970).

\bibitem{Dillamore}
I.~Dillamore, H.~Katoh, The mechanisms of recrystallization in cubic metals
  with particular reference to their orientation-dependence, Metal Science
  8~(1) (1974) 73--83.

\bibitem{samajdar1995}
I.~Samajdar, R.~D. Doherty, Role of s[\{123\} $\langle 634\rangle$]
  orientations in the preferred nucleation of cube grains in recrystallization
  of fcc metals, Scripta metallurgica et materialia 32~(6) (1995).

\bibitem{doherty1998cube}
R.~D. Doherty, L.-C. Chen, I.~Samajdar, Cube recrystallization
  texture—experimental results and modeling, Materials Science and
  Engineering: A 257~(1) (1998) 18--36.

\bibitem{samajdar1999}
I.~Samajdar, B.~Verlinden, L.~Rabet, P.~Van~Houtte, Recrystallization texture
  in a cold rolled commercial purity aluminum: on the plausible macro-and
  micro-mechanisms, Materials Science and Engineering: A 266~(1-2) (1999)
  146--154.

\bibitem{beck1952}
P.~A. Beck, H.~Hu, Annealing textures in rolled face-centered cubic metals, JOM
  4~(1) (1952) 83--90.

\bibitem{liebmann1956}
B.~Liebmann, K.~Lucke, G.~Masing, Orientation dependency of the rate of growth
  during primary recrystallization of a1 single crystals, Z. Metallkd 47 (1956)
  57--63.

\bibitem{lucke1974}
K.~L{\"u}cke, The orientation dependence of grain boundary motion and the
  formation of recrystallization textures, Canadian Metallurgical Quarterly
  13~(1) (1974) 261--274.

\bibitem{sindel1970}
M.~Sindel, G.~K{\"o}hlhoff, K.~L{\"u}cke, B.~Duggan, Development of cube
  texture in coarse grained copper, Textures and microstructures 12 (1990).

\bibitem{akef1991}
A.~Akef, J.~Driver, Orientation splitting of cube-oriented face-centred cubic
  crystals in plane strain compression, Materials Science and Engineering: A
  132 (1991) 245--255.

\bibitem{basson2000}
F.~Basson, J.~Driver, Deformation banding mechanisms during plane strain
  compression of cube-oriented fcc crystals, Acta materialia 48~(9) (2000)
  2101--2115.

\bibitem{wert1997}
J.~Wert, Q.~Liu, N.~Hansen, Dislocation boundary formation in a cold-rolled
  cube-oriented al single crystal, Acta materialia 45~(6) (1997) 2565--2576.

\bibitem{doherty1997}
R.~D. Doherty, Recrystallization and texture, Progress in materials science
  42~(1-4) (1997) 39--58.

\bibitem{liu1998}
Q.~Liu, N.~Hansen, C.~Maurice, J.~Driver, Heterogeneous microstructures and
  microtextures in cube-oriented al crystals after channel die compression,
  Metallurgical and Materials Transactions A 29~(9) (1998) 2333--2344.

\bibitem{raabe2004}
D.~Raabe, Z.~Zhao, F.~Roters, Study on the orientational stability of
  cube-oriented fcc crystals under plane strain by use of a texture component
  crystal plasticity finite element method, Scripta Materialia 50~(7) (2004)
  1085--1090.

\bibitem{samajdar1998}
I.~Samajdar, R.~Doherty, Cube recrystallization texture in warm deformed
  aluminum: understanding and prediction, Acta materialia 46~(9) (1998)
  3145--3158.

\bibitem{HIRSCH1988}
J.~Hirsch, K.~L{\"u}cke, M.~Hatherly, Overview no. 76: mechanism of deformation
  and development of rolling textures in polycrystalline fcc metals—iii. the
  influence of slip inhomogeneities and twinning, Acta Metallurgica 36~(11)
  (1988) 2905--2927.

\bibitem{jensen1995}
D.~J. Jensen, Growth rates and misorientation relationships between growing
  nuclei/grains and the surrounding deformed matrix during recrystallization,
  Acta Metallurgica et Materialia 43~(11) (1995) 4117--4129.

\bibitem{dorte}
D.~J. Jensen, Orientation aspects of growth during recrystallization, Ph.D.
  thesis, Danmarks Tekniske Universitet, Technical University of Denmark (3
  1997).

\bibitem{solberg}
E.~Nes, J.~Solberg, Growth of cube grains during recrystaiiization of
  aluminium, Materials science and technology 2~(1) (1986) 19--21.

\bibitem{vandermeer1995}
R.~Vandermeer, D.~J. Jensen, Quantifying recrystallization nucleation and
  growth kinetics of cold-worked copper by microstructural analysis,
  Metallurgical and Materials Transactions A 26~(9) (1995) 2227--2235.

\bibitem{jensen1997}
D.~J. Jensen, Simulation of recrystallization microstructures and textures:
  Effects of preferential growth, Metallurgical and Materials Transactions A
  28~(1) (1997) 15--25.

\bibitem{bernier2011}
J.~V. Bernier, N.~R. Barton, U.~Lienert, M.~P. Miller, Far-field high-energy
  diffraction microscopy: a tool for intergranular orientation and strain
  analysis, The Journal of Strain Analysis for Engineering Design 46~(7) (2011)
  527--547.

\bibitem{chatterjee2016}
K.~Chatterjee, A.~Venkataraman, T.~Garbaciak, J.~Rotella, M.~Sangid, A.~J.
  Beaudoin, P.~Kenesei, J.~Park, A.~Pilchak, Study of grain-level deformation
  and residual stresses in ti-7al under combined bending and tension using high
  energy diffraction microscopy (hedm), International Journal of Solids and
  Structures 94 (2016) 35--49.

\bibitem{wang2017}
L.~Wang, Z.~Zheng, H.~Phukan, P.~Kenesei, J.-S. Park, J.~Lind, R.~Suter, T.~R.
  Bieler, Direct measurement of critical resolved shear stress of prismatic and
  basal slip in polycrystalline ti using high energy x-ray diffraction
  microscopy, Acta Materialia 132 (2017) 598--610.

\bibitem{tayon2019}
W.~A. Tayon, K.~E. Nygren, R.~E. Crooks, D.~C. Pagan, In-situ study of planar
  slip in a commercial aluminum-lithium alloy using high energy x-ray
  diffraction microscopy, Acta Materialia 173 (2019) 231--241.

\bibitem{marthinsen}
T.~Furu, K.~Marthinsen, E.~Nes, Modelling recrystallisation, Materials science
  and technology 6~(11) (1990) 1093--1102.

\bibitem{srolovitz}
D.~Srolovitz, G.~Grest, M.~Anderson, A.~Rollett, Computer simulation of
  recrystallization—ii. heterogeneous nucleation and growth, Acta
  metallurgica 36~(8) (1988) 2115--2128.

\bibitem{raabe2000}
D.~Raabe, R.~C. Becker, Coupling of a crystal plasticity finite-element model
  with a probabilistic cellular automaton for simulating primary static
  recrystallization in aluminium, Modelling and Simulation in Materials Science
  and Engineering 8~(4) (2000) 445.

\bibitem{takaki}
T.~Takaki, A.~Yamanaka, Y.~Tomita, Phase-field modeling and simulation of
  nucleation and growth of recrystallized grains, in: Materials Science Forum,
  Vol. 558, Trans Tech Publ, 2007, pp. 1195--1200.

\bibitem{abrivard2012}
G.~Abrivard, E.~P. Busso, S.~Forest, B.~Appolaire, Phase field modelling of
  grain boundary motion driven by curvature and stored energy gradients. part
  ii: application to recrystallisation, Philosophical magazine 92~(28-30)
  (2012) 3643--3664.

\bibitem{moelans1}
N.~Moelans, A.~Godfrey, Y.~Zhang, D.~J. Jensen, Phase-field simulation study of
  the migration of recrystallization boundaries, Physical Review B 88~(5)
  (2013) 054103.

\bibitem{chenLQ}
L.~Chen, J.~Chen, R.~Lebensohn, Y.~Ji, T.~Heo, S.~Bhattacharyya, K.~Chang,
  S.~Mathaudhu, Z.~Liu, L.-Q. Chen, An integrated fast fourier transform-based
  phase-field and crystal plasticity approach to model recrystallization of
  three dimensional polycrystals, Computer Methods in Applied Mechanics and
  Engineering 285 (2015) 829--848.

\bibitem{alvi2008}
M.~H. Alvi, S.~Cheong, J.~Suni, H.~Weiland, A.~Rollett, Cube texture in
  hot-rolled aluminum alloy 1050 (aa1050)—nucleation and growth behavior,
  Acta materialia 56~(13) (2008) 3098--3108.

\bibitem{brahme2008}
A.~Brahme, J.~Fridy, H.~Weiland, A.~D. Rollett, Modeling texture evolution
  during recrystallization in aluminum, Modelling and Simulation in Materials
  Science and Engineering 17~(1) (2008) 015005.

\bibitem{field}
K.~Adam, D.~Z{\"o}llner, D.~P. Field, 3d microstructural evolution of primary
  recrystallization and grain growth in cold rolled single-phase aluminum
  alloys, Modelling and Simulation in Materials Science and Engineering 26~(3)
  (2018) 035011.

\bibitem{zecevic2019}
M.~Zecevic, R.~A. Lebensohn, R.~J. McCabe, M.~Knezevic, Modelling
  recrystallization textures driven by intragranular fluctuations implemented
  in the viscoplastic self-consistent formulation, Acta Materialia 164 (2019)
  530--546.

\bibitem{moulinec}
H.~Moulinec, P.~Suquet, A numerical method for computing the overall response
  of nonlinear composites with complex microstructure, Computer methods in
  applied mechanics and engineering 157~(1-2) (1998) 69--94.

\bibitem{lebensohn1}
R.~A. Lebensohn, A.~K. Kanjarla, P.~Eisenlohr, An elasto-viscoplastic
  formulation based on fast fourier transforms for the prediction of
  micromechanical fields in polycrystalline materials, International Journal of
  Plasticity 32 (2012) 59--69.

\bibitem{arul1}
M.~A. Kumar, A.~Kanjarla, S.~Niezgoda, R.~Lebensohn, C.~Tom{\'e}, Numerical
  study of the stress state of a deformation twin in magnesium, Acta Materialia
  84 (2015) 349--358.

\bibitem{arul2}
M.~A. Kumar, I.~Beyerlein, R.~Lebensohn, C.~Tome, Modeling the effect of
  neighboring grains on twin growth in hcp polycrystals, Modelling and
  Simulation in Materials Science and Engineering 25~(6) (2017) 064007.

\bibitem{tasan}
C.~C. Tasan, M.~Diehl, D.~Yan, C.~Zambaldi, P.~Shanthraj, F.~Roters, D.~Raabe,
  Integrated experimental--simulation analysis of stress and strain
  partitioning in multiphase alloys, Acta Materialia 81 (2014) 386--400.

\bibitem{diehl2017}
M.~Diehl, D.~An, P.~Shanthraj, S.~Zaefferer, F.~Roters, D.~Raabe, Crystal
  plasticity study on stress and strain partitioning in a measured 3d dual
  phase steel microstructure, Physical Mesomechanics 20~(3) (2017) 311--323.

\bibitem{lebensohn3}
R.~A. Lebensohn, R.~Brenner, O.~Castelnau, A.~D. Rollett, Orientation
  image-based micromechanical modelling of subgrain texture evolution in
  polycrystalline copper, Acta Materialia 56~(15) (2008) 3914--3926.

\bibitem{connor}
C.~Slone, S.~Chakraborty, J.~Miao, E.~P. George, M.~J. Mills, S.~Niezgoda,
  Influence of deformation induced nanoscale twinning and fcc-hcp
  transformation on hardening and texture development in medium-entropy crconi
  alloy, Acta Materialia 158 (2018) 38--52.

\bibitem{zhao1}
P.~Zhao, T.~S.~E. Low, Y.~Wang, S.~R. Niezgoda, An integrated full-field model
  of concurrent plastic deformation and microstructure evolution: application
  to 3d simulation of dynamic recrystallization in polycrystalline copper,
  International Journal of Plasticity 80 (2016) 38--55.

\bibitem{zhao2}
P.~Zhao, Y.~Wang, S.~R. Niezgoda, Microstructural and micromechanical evolution
  during dynamic recrystallization, International Journal of Plasticity 100
  (2018) 52--68.

\bibitem{lebensohn_rollett}
R.~A. Lebensohn, A.~D. Rollett, Spectral methods for full-field micromechanical
  modelling of polycrystalline materials, Computational Materials Science 173
  (2020) 109336.

\bibitem{ma1}
A.~Ma, F.~Roters, A constitutive model for fcc single crystals based on
  dislocation densities and its application to uniaxial compression of
  aluminium single crystals, Acta materialia 52~(12) (2004) 3603--3612.

\bibitem{ma2}
A.~Ma, F.~Roters, D.~Raabe, A dislocation density based constitutive model for
  crystal plasticity fem including geometrically necessary dislocations, Acta
  Materialia 54~(8) (2006) 2169--2179.

\bibitem{conrad}
H.~Conrad, Thermally activated deformation of metals, JOM 16~(7) (1964)
  582--588.

\bibitem{kocks}
U.~Kocks, Laws for work-hardening and low-temperature creep, Journal of
  engineering materials and technology 98~(1) (1976) 76--85.

\bibitem{mughrabi}
U.~Essmann, H.~Mughrabi, Annihilation of dislocations during tensile and cyclic
  deformation and limits of dislocation densities, Philosophical Magazine A
  40~(6) (1979) 731--756.

\bibitem{kords}
C.~Kords, On the role of dislocation transport in the constitutive description
  of crystal plasticity, epubli, 2013.

\bibitem{dai}
H.~Dai, Geometrically-necessary dislocation density in continuum plasticity
  theory, fem implementation and applications, Ph.D. thesis, Massachusetts
  Institute of Technology (1997).

\bibitem{arsenlis2}
A.~Arsenlis, D.~Parks, Crystallographic aspects of geometrically-necessary and
  statistically-stored dislocation density, Acta materialia 47~(5) (1999)
  1597--1611.

\bibitem{bulatov}
V.~Bulatov, F.~F. Abraham, L.~Kubin, B.~Devincre, S.~Yip, Connecting atomistic
  and mesoscale simulations of crystal plasticity, Nature 391~(6668) (1998)
  669--672.

\bibitem{kubin2008}
L.~Kubin, B.~Devincre, T.~Hoc, Modeling dislocation storage rates and mean free
  paths in face-centered cubic crystals, Acta Materialia 56~(20) (2008)
  6040--6049.

\bibitem{madec2017}
R.~Madec, L.~P. Kubin, Dislocation strengthening in fcc metals and in bcc
  metals at high temperatures, Acta Materialia 126 (2017) 166--173.

\bibitem{martinez2008}
E.~Martinez, J.~Marian, A.~Arsenlis, M.~Victoria, J.~M. Perlado, Atomistically
  informed dislocation dynamics in fcc crystals, Journal of the Mechanics and
  Physics of Solids 56~(3) (2008) 869--895.

\bibitem{niezgoda}
S.~R. Niezgoda, A.~K. Kanjarla, I.~J. Beyerlein, C.~N. Tom{\'e}, Stochastic
  modeling of twin nucleation in polycrystals: an application in hexagonal
  close-packed metals, International journal of plasticity 56 (2014) 119--138.

\bibitem{doherty1997current}
R.~Doherty, D.~Hughes, F.~Humphreys, J.~J. Jonas, D.~J. Jensen, M.~Kassner,
  W.~King, T.~McNelley, H.~McQueen, A.~Rollett, Current issues in
  recrystallization: a review, Materials Science and Engineering: A 238~(2)
  (1997) 219--274.

\bibitem{szpunar}
N.~Rajmohan, J.~Szpunar, A new model for recrystallization of heavily
  cold-rolled aluminum using orientation-dependent stored energy, Acta
  materialia 48~(13) (2000) 3327--3340.

\bibitem{holm2003}
E.~A. Holm, M.~A. Miodownik, A.~D. Rollett, On abnormal subgrain growth and the
  origin of recrystallization nuclei, Acta Materialia 51~(9) (2003) 2701--2716.

\bibitem{rollett}
S.~Wang, E.~A. Holm, J.~Suni, M.~H. Alvi, P.~N. Kalu, A.~D. Rollett, Modeling
  the recrystallized grain size in single phase materials, Acta Materialia
  59~(10) (2011) 3872--3882.

\bibitem{chenLQ1}
L.-Q. Chen, Y.~Wang, The continuum field approach to modeling microstructural
  evolution, Jom 48~(12) (1996) 13--18.

\bibitem{chenLQ2}
D.~Fan, L.-Q. Chen, Computer simulation of grain growth using a continuum field
  model, Acta Materialia 45~(2) (1997) 611--622.

\bibitem{moelans2008}
N.~Moelans, B.~Blanpain, P.~Wollants, Quantitative analysis of grain boundary
  properties in a generalized phase field model for grain growth in anisotropic
  systems, Physical Review B 78~(2) (2008) 024113.

\bibitem{moelans3}
N.~Moelans, A quantitative and thermodynamically consistent phase-field
  interpolation function for multi-phase systems, Acta Materialia 59~(3) (2011)
  1077--1086.

\bibitem{bachmann2010}
F.~Bachmann, R.~Hielscher, H.~Schaeben, Texture analysis with mtex--free and
  open source software toolbox, in: Solid State Phenomena, Vol. 160, Trans Tech
  Publ, 2010, pp. 63--68.

\bibitem{bronkhorst}
C.~A. Bronkhorst, Plastic deformation and crystallographic texture evolution in
  face-centered cubic metals, Ph.D. thesis, Massachusetts Institute of
  Technology (1991).

\bibitem{kallend}
J.~Kallend, Y.~Huang, Orientation dependence of stored energy of cold work in
  50\% cold rolled copper, Metal science 18~(7) (1984) 381--386.

\bibitem{godfrey}
A.~Godfrey, N.~Hansen, D.~J. Jensen, Microstructural-based measurement of local
  stored energy variations in deformed metals, Metallurgical and Materials
  Transactions A 38~(13) (2007) 2329--2339.

\bibitem{vandermeer}
R.~Vandermeer, D.~J. Jensen, E.~Woldt, Grain boundary mobility during
  recrystallization of copper, Metallurgical and Materials Transactions A
  28~(3) (1997) 749--754.

\bibitem{bacroix}
G.~Mohamed, B.~Bacroix, Role of stored energy in static recrystallization of
  cold rolled copper single and multicrystals, Acta materialia 48~(13) (2000)
  3295--3302.

\bibitem{doherty1985}
R.~D. Doherty, Nucleation and growth kinetics of different recrystallization
  texture components, Scripta metallurgica 19~(8) (1985) 927--930.

\bibitem{mackenzie}
J.~Mackenzie, Second paper on statistics associated with the random
  disorientation of cubes, Biometrika 45~(1-2) (1958) 229--240.

\bibitem{albou2010}
A.~Albou, S.~Raveendra, P.~Karajagikar, I.~Samajdar, C.~Maurice, J.~H. Driver,
  Direct correlation of deformation microstructures and cube recrystallization
  nucleation in aluminium, Scripta Materialia 62~(7) (2010) 469--472.

\bibitem{lauridsen2003}
E.~Lauridsen, H.~Poulsen, S.~Nielsen, D.~J. Jensen, Recrystallization kinetics
  of individual bulk grains in 90\% cold-rolled aluminium, Acta Materialia
  51~(15) (2003) 4423--4435.

\bibitem{lin}
F.~Lin, Y.~Zhang, W.~Pantleon, D.~Juul~Jensen, Supercube grains leading to a
  strong cube texture and a broad grain size distribution after
  recrystallization, Philosophical Magazine 95~(22) (2015) 2427--2449.

\bibitem{kazaryan1}
A.~Kazaryan, Y.~Wang, S.~A. Dregia, B.~R. Patton, Generalized phase-field model
  for computer simulation of grain growth in anisotropic systems, Phys. Rev. B
  61 (2000) 14275--14278.

\bibitem{kazaryan2}
A.~Kazaryan, Y.~Wang, S.~Dregia, B.~Patton, Grain growth in anisotropic
  systems: comparison of effects of energy and mobility, Acta Materialia
  50~(10) (2002) 2491--2502.

\end{thebibliography}

\end{document}